\input harvmac.tex

\def\IB{\relax\hbox{$\inbar\kern-.3em{\rm B}$}}

\def\IC{\relax\hbox{$\inbar\kern-.3em{\rm C}$}}

\def\ID{\relax\hbox{$\inbar\kern-.3em{\rm D}$}}

\def\IE{\relax\hbox{$\inbar\kern-.3em{\rm E}$}}

\def\IF{\relax\hbox{$\inbar\kern-.3em{\rm F}$}}

\def\IG{\relax\hbox{$\inbar\kern-.3em{\rm G}$}}

\def\IGa{\relax\hbox{${\rm I}\kern-.18em\Gamma$}}

\def\IH{\relax{\rm I\kern-.18em H}}

\def\IK{\relax{\rm I\kern-.18em K}}

\def\IL{\relax{\rm I\kern-.18em L}}

\def\IP{\relax{\rm I\kern-.18em P}}

\def\IR{\relax{\rm I\kern-.18em R}}

\def\IZ{\relax\ifmmode\mathchoice    

{\hbox{\cmss Z\kern-.4em Z}}{\hbox{\cmss Z\kern-.4em Z}}
{\lower.9pt\hbox{\cmsss Z\kern-.4em Z}}
{\lower1.2pt\hbox{\cmsss Z\kern-.4em Z}}\else{\cmss Z\kern-.4em Z}\fi}


\def\CF {{\cal F}}

\def\CL {{\cal L}}




\def\Tr{\rm Tr}



\def\unlockat{\catcode`\@=11}

\def\lockat{\catcode`\@=12}

\unlockat


\def\newsec#1{\global\advance\secno by1\message{(\the\secno. #1)}
\global\subsecno=0\global\subsubsecno=0\eqnres@t\noindent
{\bf\the\secno. #1}
\writetoca{{\secsym} {#1}}\par\nobreak\medskip\nobreak}

\global\newcount\subsecno \global\subsecno=0
\def\subsec#1{\global\advance\subsecno
by1\message{(\secsym\the\subsecno. #1)}
\ifnum\lastpenalty>9000\else\bigbreak\fi\global\subsubsecno=0
\noindent{\it\secsym\the\subsecno. #1}
\writetoca{\string\quad {\secsym\the\subsecno.} {#1}}
\par\nobreak\medskip\nobreak}
\global\newcount\subsubsecno \global\subsubsecno=0
\def\subsubsec#1{\global\advance\subsubsecno by1
\message{(\secsym\the\subsecno.\the\subsubsecno. #1)}
\ifnum\lastpenalty>9000\else\bigbreak\fi
\noindent\quad{\secsym\the\subsecno.\the\subsubsecno.}{#1}
\writetoca{\string\qquad{\secsym\the\subsecno.\the\subsubsecno.}{#1}}
\par\nobreak\medskip\nobreak}

\def\subsubseclab#1{\DefWarn#1\xdef
#1{\noexpand\hyperref{}{subsubsection}%
{\secsym\the\subsecno.\the\subsubsecno}%
{\secsym\the\subsecno.\the\subsubsecno}}%
\writedef{#1\leftbracket#1}\wrlabeL{#1=#1}}
\lockat

\def\IL{\relax{\rm I\kern-.18em L}}

\def\IH{\relax{\rm I\kern-.18em H}}

\def\IR{\relax{\rm I\kern-.18em R}}


\def\dbend{\lower3.5pt\hbox{\manual\char127}}

\def\c{\cdot}
\def\IZ{\relax\ifmmode\mathchoice
{\hbox{\cmss Z\kern-.4em Z}}{\hbox{\cmss Z\kern-.4em Z}}
{\lower.9pt\hbox{\cmsss Z\kern-.4em Z}}
{\lower1.2pt\hbox{\cmsss Z\kern-.4em Z}}\else{\cmss Z\kern-.4em
Z}\fi}


\def\IZ{\relax\ifmmode\mathchoice
{\hbox{\cmss Z\kern-.4em Z}}{\hbox{\cmss Z\kern-.4em Z}}
{\lower.9pt\hbox{\cmsss Z\kern-.4em Z}}
{\lower1.2pt\hbox{\cmsss Z\kern-.4em Z}}\else{\cmss Z\kern-.4em
Z}\fi}

\def\IB{\relax{\rm I\kern-.18em B}}

\def\IC{{\relax\hbox{$\inbar\kern-.3em{\rm C}$}}}

\def\ID{\relax{\rm I\kern-.18em D}}

\def\IE{\relax{\rm I\kern-.18em E}}

\def\IF{\relax{\rm I\kern-.18em F}}

\def\IG{\relax\hbox{$\inbar\kern-.3em{\rm G}$}}

\def\IGa{\relax\hbox{${\rm I}\kern-.18em\Gamma$}}

\def\IH{\relax{\rm I\kern-.18em H}}

\def\II{\relax{\rm I\kern-.18em I}}

\def\IK{\relax{\rm I\kern-.18em K}}

\def\IP{\relax{\rm I\kern-.18em P}}

\def\inbar{\,\vrule height1.5ex width.4pt depth0pt}

\font\cmss=cmss10 \font\cmsss=cmss10 at 7pt

\def\IR{\relax{\rm I\kern-.18em R}}

\def\Tr{\rm Tr}


\def\boxit#1{\vbox{\hrule\hbox{\vrule\kern8pt
\vbox{\hbox{\kern8pt}\hbox{\vbox{#1}}\hbox{\kern8pt}}
\kern8pt\vrule}\hrule}}
\def\mathboxit#1{\vbox{\hrule\hbox{\vrule\kern8pt\vbox{\kern8pt
\hbox{$\displaystyle #1$}\kern8pt}\kern8pt\vrule}\hrule}}


\def\inbar{\,\vrule height1.5ex width.4pt depth0pt}

\font\cmss=cmss10 \font\cmsss=cmss10 at 7pt
\def\IR{\relax{\rm I\kern-.18em R}}

\def\Tr{\rm Tr}


%

\def\hh{hep-th/}

\lref\simons{  J. Cheeger and J. Simons, {\it Differential Characters
and
Geometric Invariants},
 Stony Brook Preprint, (1973), unpublished.}

\lref\cargese{ L.~Baulieu,
{Algebraic Quantization of Gauge Theories}, Lectures given
at the Carg\`ese Summer School on   ``Perspectives in
fields and particles'', 198, p1, Proc. ed. by J.L.
 Basdevant and M. Levy  (Plenum Press, New-York, 1983).}

 \lref\antifields{   L. Baulieu, M. Bellon, S. Ouvry and  C.
Wallet,   Phys.Letters   B252  (1990) 387;  M.  Bocchichio, Phys.
Lett.    B187     (1987) 322 and    B 192  (1987) 31; R.  Thorn,  
  Nucl. Phys.   B257 
(1987) 61. }

 \lref\thompson{  G. Thompson,  Annals Phys. 205 (1991) 130; 
  J.M.F. Labastida and  M. Pernici, Phys. Lett. 212B  (1988) 56; 
  D. Birmingham, M.Blau,  M. Rakowski and G. Thompson,  Phys. Rep. 209
(1991) 129.}

\lref\Acharya{ B. Acharya, M. O'Loughlin and
B. Spence, {\it  Higher Dimensional Analogues of Donaldson-Witten Theory}, 
hepth/9705138, Nucl. Phys. B503 (1997) 657-674.}

  \lref\tonin{ Tonin}

 \lref\seibergsix{  O. Aharony, M. Berkooz, N. Seiberg,  {\it Light-Cone
Description of (2,0) Superconformal Theories in Six Dimensions},
  hep-th/9712117\semi  O.  J.  Ganor, David R.  Morrison, N.  Seiberg,
 {\it
Branes, Calabi-Yau Spaces, and Toroidal Compactification of the N=1
Six-Dimensional $E_8$ Theory}, hep-th/9610251, Nucl. Phys.  B487 (1997)
93-127\semi
N.  Seiberg,
{\it Non-trivial Fixed Points of The Renormalization Group in Six
Dimensions},  hep-th/9609161, Phys. Lett.  B390 (1997) 169-171. 
}

 \lref\sixseiberg{
N. Seiberg, {\it Notes on Theories with 16 Supercharges},
hep-th/9705117,  Nucl.Phys.Proc.Suppl. 67 (1998) 158-171}

 \lref\lerche{See, e.g., W. Lerche,
{\it
Elliptic Index and Superstring Effective Actions}, Nucl. Phys.  B308
(1988)
101-126, p.109, eq. 2.15. }

\lref\warner{ W. Lerche and  S. Stieberger
{\it Prepotential, Mirror Map and F-Theory on K3},
hep-th/9804176 ;
C. Bachas, C. Fabre, E. Kiritsis, N. A. Obers and P. Vanhove, {\it Heterotic  type I duality and D-brane instantons}, hep-th/9707126, Nucl.Phys. B509 (1998) 33-52; W. Lerche, N.Warner, and S. Stieberger, in preparation.}

\lref\wittentopo { E.  Witten,  {\it  Topological Quantum Field Theory},
\hh9403195,
Commun.  Math. Phys.  {117} (1988) 353.  }

\lref\wittentwist { E.  Witten, {\it Supersymmetric Yang--Mills theory
on a
four-manifold}, J.  Math.  Phys.  {35} (1994) 5101.}

\lref\wittensix{E.  Witten, {\it New  Gauge  Theories In Six
Dimensions},
  hep-th/9710065. }

\lref\wittendual {E.  Witten,
 {\it On Self-Duality in Abelian Gauge Theories,}   hep-th/9505186. }

\lref\sw {N. Seiberg and E.  Witten, 
 {\it Monopoles, Duality and Chiral Symmetry Breaking in N=2
Supersymmetric QCD,}, hep-th/9408099,Nucl.Phys. B431 (1994) 484-550\semi
 { \it Monopole Condensation  and Confinement in $N=2$ Supersymmetric
Yang-Mills Theory,}  hep-th/9407087, Nucl.Phys. B426 (1994) 19-52;  
B430 (1994) 485-486.
}


\lref\bfss{T.~Banks, W.Fischler, S.H.~Shenker, L.~Susskind, {\it M Theory
as a Matrix Model : A Conjecture}, \hh9610043.}
\lref\seiberg{N.~ Seiberg
``Why is the Matrix Model Correct?'', 
hep-th/9710009, 
Phys.Rev.Lett. 79 (1997) 3577-3580}
\lref\sen{A.~ Sen, ``D0 Branes on $T^n$ and Matrix Theory'', 
hep-th/9709220}
\lref\bsv{M. Bershadsky, V. Sadov, C. Vafa,
``D-Branes and Topological Field Theories'', hep-th/9511222 
Nucl.Phys. {\bf B}463 (1996) 420-434}
\lref\vafapuzz{C.~Vafa, ``Puzzles at Large N'', hep-th/9804172}
\lref\DVV{R.~Dijkgraaf, E.~Verlinde, H.~Verlinde, {\it Matrix String
Theory}, \hh9703030.}
\lref\ikkt{N.~Ishibashi, H.~Kawai, Y.~Kitazawa, A.~Tsuchiya,
``A Large-N Reduced Model as Superstring'',  hep-th/9612115, 
Nucl.~Phys. {\bf B}498 (1997) 467}
\lref\ss{S.~ Sethi, M.~ Stern, 
``D-Brane Bound States Redux'', hep-th/9705046,
Commun.Math.Phys. 194 (1998) 675-705}
\lref\mns{G.~Moore, N.~Nekrasov, S.~Shatashvili,
``D-particle bound states and generalized instantons'', hep-th/9803265 }
 

\lref\orlando{
 O. Alvarez, L. A. Ferreira and J. Sanchez Guillen, 
 {\it  A New Approach to Integrable Theories in any Dimension},
hep-th/9710147.
} 
\lref\bks{  L.~Baulieu, H.~Kanno and I.~Singer,
{\it Special Quantum Field Theories in Eight and Other Dimensions},
hep-th/9704167, Talk given at
APCTP Winter School on Dualities in String Theory  (Sokcho, Corea),
February 24-28, 1997\semi
  L.~Baulieu, H.~Kanno and I.~Singer, {\it Cohomological Yang--Mills
Theory
in Eight Dimensions}, hep-th/9705127, to appear in Commun. Math. Phys. }

\lref\west{  L.~Baulieu and P.~West,
{  \it Six-Dimensional TQFTs and Twisted Supersymmetry},
 hep-th/9805200\semi
L.~Baulieu and E. Rabinovici, {\it Self-Duality and New TQFTs for Forms,
 }
hep-th/9805122.}

\lref\bv{ I.A. Batalin  and V.A. Vilkowisky,    Phys. Rev.  
D28  (1983) 2567\semi  M. Henneaux,  Phys. Rep.  126   (1985)
1;            
M. Henneaux and C. Teitelboim, {\it Quantization of Gauge Systems,}
Princeton University Press,  Princeton (1992).}

\lref\bfss{T.~Banks, W.Fischler, S.H.~Shenker, L.~Susskind, {\it M Theory
as a Matrix Model : A Conjecture}, \hh9610043.}

\lref\bs { L. Baulieu and I. M. Singer, {\it Topological Yang--Mills
Symmetry}, Nucl. Phys. Proc. Suppl.  
15B (1988) 12\semi  L. Baulieu, {\it On the Symmetries of Topological
Quantum
Field Theories},   hep-th/ 9504015, Int. J. Mod. Phys. A10 (1995)
4483\semi S. Cordes, G. Moore and S. Ramgoolam , {\it Lectures on 2D Yang-Mills
theory,
Equivariant Cohomology and Topological String Theory}, Lectures poresented at
1994 Les Houches Summer School and at Trieste 1994 Spring School on
Superstrings, hep-th/9411210, http://xxx.lanl.gov/lh94 \semi R. Dijkgraaf and G. Moore,
 {\it Balanced Topological Field Theories},
hep-th/9608169,   Commun. Math. Phys. 185 (1997) 411.}

\lref\kyoto {  L. Baulieu,   {\it Field Antifield Duality, p-Form Gauge
Fields and Topological Quantum
Field Theories},     hep-th/9512026,  Nucl. Phys.  B478 (1996) 431.  }

\lref\geneva {  L. Baulieu,   {\it On Forms with Non-Abelian Charges and
Their Dualities},     hep-th/9808055, to appear in Phys. Lett. B.  }

\lref\sezgin {
 L. Baulieu, E. Bergshoeff and E. Sezgin,
 {\it
Open BRST Algebra, Ghost Unification and String Field Theory, } 
Nucl. Phys.   B307  (1988) 348.  }

\lref\strings {  L.  Baulieu, M. B. Green and E. Rabinovici {\it A
Unifying
Topological Action for Heterotic and  Type II Superstring  Theories},
hep-th/9606080, Phys.Lett. B386 (1996) 91\semi
{\it   Superstrings from   Theories with $N>1$ World Sheet
Supersymmetry},
 hep-th/9611136, Nucl. Phys. B498 (1997). }

\lref\sourlas{  G. Parisi and N. Sourlas,
{\it Random Magnetic Fields, Supersymmetry and Negative Dimensions}, 
Phys.
Rev. Lett.  43 (1979) 744; Nucl.  Phys.  B206 (1982) 321.  }

\lref\SalamSezgin{A.  Salam  and  E.  Sezgin,
{\it Supergravities in diverse dimensions}, vol.  1, p. 119\semi
P.  Howe, G.  Sierra and P.  Townsend, Nucl Phys B221 (1983) 331.}

\lref\nekrasov{ A. Losev, G. Moore, N. Nekrasov, S. Shatashvili,
{\it
Four-Dimensional Avatars of Two-Dimensional RCFT},  hep-th/9509151,
Nucl.  Phys.  Proc.  Suppl.   46 (1996) 130\semi L.  Baulieu, A.  Losev,
N. 
Nekrasov  {\it Chern-Simons and Twisted Supersymmetry in Higher
Dimensions},  hep-th/9707174, to appear in Nucl.  Phys.  B.  }

\lref\mw{G. Moore and E. Witten, {\it 
Integration over the u-plane in Donaldson theory
}, hep-th/9709193}

\lref\lns{A. Losev, N. Nekrasov and S. Shatashvili,
{\it  Issues in Topological Gauge Theory},
 hep-th/9711108,  Nucl. Phys. B534, 549, 1998;
{\it Testing Seiberg-Witten Equations},
hep-th/9801061,
Contribution  to NATO Advanced Study Institute on Strings, Branes and Dualities, Cargese, France, 1997.
}

\lref\mm{ 
 M. Marino and  G.  Moore, {\it Donaldson invariants for nonsimply connected manifolds},
hep-th/9804104;
{\it The Donaldson-Witten function for gauge groups of rank larger than one},
 hep-th/9802185;
{\it Integrating over the Coulomb branch in N=2 gauge theory},
hep-th/9712062, Nucl.Phys.Proc.Suppl. 68 (1998) 336-347.
}

\Title{ \vbox{\baselineskip12pt\hbox{hep-th/9811198}
\hbox{CERN/TH-98-258}
 \hbox{Yale/YCTP-P28-98}
\hbox{LPTHE/98-59 }}}
{\vbox{\centerline{       DUALITY FROM TOPOLOGICAL SYMMETRY }}}

\medskip

\centerline{Laurent Baulieu $^{a, b}$ \footnote{*}{UMR-CNRS 
associ\'ee aux  Universit\'es Pierre et
Marie Curie (Paris VI) et Denis Diderot (Paris~VII)}and
Samson L. Shatashvili $^{a, c}$\footnote{**}{On leave of absence from 
St. Petersburg Steklov
Mathematical Institute.}}   

\vskip 0.5cm

\centerline{$^{a}$ CERN, 
Geneva, Switzerland }
\centerline{$^{b}$ LPTHE, Paris, France}
\centerline{$^{c}$ Department of Physics, Yale University,
 Box 208120, New Haven, CT 006520 USA}

\vskip 0.1cm

 \medskip

\vskip  1cm

\noindent

We describe   topological gauge theories for which 
duality properties are  encoded by construction.
We study them for compact manifolds of dimensions 
four, eight and two.
The fields and their duals are treated symmetrically, within the 
context of field--antifield unification.  Dual formulations 
correspond to different gauge-fixings of the  topological
symmetry. We also
describe 
    novel features in eight-dimensional theories,  and speculate about
their possible   ``Abelian" descriptions.


\Date{August  1998}

\def\e{\epsilon}

\def\demi{{1\over 2}}

\def\pa{\partial}

\def\a{\alpha}

\def\b{\beta}

\def\c{\gamma}

\def\m{\mu}

\def\n{\nu}

\def\r{\rho}

\def\L{L}

\def\X{X}

\def\V{V}

\def\P{\Psi}

\def\F{\Phi}

\def\d{\delta}

\newsec{Introduction}

The   relationship  between
topological gauge theories
 for
four-manifold
invariants and     physical supersymmetric theories  
    has attracted a lot of   interest. A
breakthrough   has occurred through the    work of Seiberg
and Witten on an exact low-energy description
of the physical $D=4, N=2$ supersymmetric Yang--Mills  theory  \sw.
 This theory
can be  topologically twisted in the weak
coupling
ultraviolet  regime to produce  the integral over instanton
moduli space of certain
characteristic classes for the correlation function of topological
observables. This  
  provides
a description of four-manifold invariants, although
a precise
compactification
procedure of the  moduli space instantons is not known in each case. By going to the
infrared 
description
of the same theory, which   is  an Abelian gauge theory with exact
electromagnetic duality,    integrals
over the 
instanton moduli space are found to  equate  integrals over the  moduli
space  of vacua
\wittendual  \mw 
\lns  \mm. The generating
function of topological correlators can be written in terms of
Seiberg--Witten monopole invariants and of the      deformed
holomorphic prepotential $\CF(a,t)$. The function 
$\CF(a,0)$  has a   description in terms
of the 
(holomorphic) symplectic geometry on $C^{2r}$, where $r$ is the  rank of
a gauge
group.  Actually,  $\CF(a,0)$   is a generating function of
$\Gamma$-invariant submanifold in
$C^{2r}$, where 
$\Gamma$ is  the finite modular subgroup (electro-magnetic duality
group) of $SP(2r,Z)$. This submanifold is described by a function 
${\CF}$ in the following
way. Denote by $\omega$ the holomorphic symplectic form on $C^{2r}$ 
and take  $\theta =
d^{-1}
\omega$. Then the restriction of
$\theta$ on the Lagrangian submanifold is defined by ${\CF}$ such that
$\theta {  |_{\CL}}= 
d{\CF}$.
The $\beta$-function of the  $N=2$ gauge theory
provides the asymptotic behavior of the prepotential, and    
the positivity of ${\rm Im}\tau$ is required, with $\tau=\partial^2
\CF$. The prepotential
$\CF(a,t)$, which describes
the correlators of observables from $H^*(X)$, is defined as a solution of
some Hamiltonian evolution of $\CF(a,0)$,  which is then a deformation
of the  Lagrangian submanifold.

In this paper we will demonstrate that the language 
of TQFTs   \bs \bks\
is adequate to describe the properties related to the duality symmetry, 
in a way that  extends refs. \kyoto  \geneva.
It provides the symmetric treatment of fields and their duals,
and we will see how   different formulations, and/or dual  pictures, can
appear in the process of different gauge-fixings of the  enlarged
topological symmetry that act on all fields. The doubling of fields,
which allows    various  
formulations,    can be generalized to other theories, in dimensions 
other than four, and for     gauge 
fields that can be forms of 
 degrees    different than  $1$. The   
  naturalness of this doubling is generally dictated by the principle
 of kinematical ghost
and gauge field  unification.  We will actually  study   topological  
theories in two and  eight dimensions, for which the approach developed
for  the four-dimensional case turns out to be generalizable and useful.
Moreover, dimensional reduction can be applied to our results,   the
most interesting case being, in our opinion,  the Yang--Mills TQFT in
eight dimensions. 

For   eight-manifolds with $Spin(7)$ holonomy,
the moduli space of four-dimensional 
instantons is replaced by that
 of octonionic
instantons, which, unfortunately,   is yet 
widely unexplored.  The eight-dimensional TQFT that was constructed in
\bks\Acharya\ is, however, quite an   attractive theory. In particular, its
dependence on an invariant   self-dual four-form 
$\Omega _4$  makes it  very interesting, since, from a physical point
of view,  such four-forms could be related to propagating four-forms in
ten dimensions. The untwisted version of this TQFT  is
nothing else  than  the ordinary
eight-dimensional  Yang--Mills supersymmetric theory, i.e.,  the 
dimensional reduction to eight dimensions   of the
$N=1,D=10$ super-Yang--Mills theory \bks.
Such a theory   also   determines the matrix string in the
light-cone gauge \bfss\DVV, after dimensional reduction to two dimensions,
and a certain form of the Seiberg--Witten theory in four dimensions.

Our treatment will  clarify   already known observations for the 
four-dimensional case. 
In the eight-dimensional case, it will  give  
interesting new features.
The   eight-dimensional  Yang--Mills theory is not renormalizable as
such.    Thus, some of the    arguments used in four dimensions 
cannot be applied directly.   Moreover, the genuine physical 
Yang--Mills   
theory on $R^8$ is infrared-free, and it   might  be  concluded  that
  non-trivial statements using
infrared  description cannot be made \foot {Notice that
the eight-dimensional Yang--Mills theory    is not fully topological,
since 
it is only independent of the metric
variations which do not change the $Spin(7)$ structure.}.

We will speculate   on   
possible resolutions of these questions.
First, we will show the possibility of a dual formulation in
eight dimensions. 
It relies on  the existence of a $Spin(7)$-invariant four-form  
$\Omega_4$, which is already known to permit the construction of 
a Yang--Mills TQFT in eight dimensions  \bks. (The existence of
$\Omega_4$ amounts to that of    octonionic structure coefficients 
in a local description.)
Here, we will show that, thanks to $\Omega_4$, duality can   
be established  
  between  a pair of  one-forms in eight dimension, and one  finally 
obtains a duality in eight dimensions, which is  parallel to that of
four dimensions. The argument will be that the  dual in eight
dimensions of a Yang--Mills field $A$, which  is a
five-form gauge field   $A_5$, can be restricted to 
a      dual one-form
$A_D$, by a partial gauge-fixing of the topological gauge symmetry,
 which
amounts   to  ``divide"    $A_5$ by $\Omega_4$,
$A_5=\Omega_4 A_D$, as well as some of the ghosts.

Secondly, we will see that, by handling 
the duality of supersymmetric theories in 
the context of TQFTs, we can  accommodate the existence 
of higher-order  interactions in the 
Yang--Mills curvatures.

  A natural way of thinking about such higher-order  
corrections, which add to the Lagrangian operators like 
quartic powers of the Yang--Mills curvature,  is the
string theory compactification down to eight dimensions.
The latter, 
  permits the replacement   of 
the ultraviolet  cut-off of the eight-dimensional  theory   by
the  string tension parameter $\alpha' $. This certainly modifies the
infrared properties  and gives a physical content to the
eight-dimensional TQFT.

A novel feature of our work is that we  will introduce  the 
${\cal O}(\a')$
 quartic interactions  (and possibly the higher-order ones, if we were
to consider 
 higher-order $\alpha'$ corrections),
by  using      $ dA+ {\cal O}(\a')t_8 dAdAdA $ instead of the
genuine   
Yang--Mills two-form curvature $dA $     
 in the octonionic gauge function of the Yang--Mills  TQFT. Here, $t_8$
is an  invariant $SO(8)$ tensor with eight indices, 
which is  is proportional  to the trace of a  
product of $\gamma$ matrices.

In the way we will proceed, the  complete   string-corrected
eight-dimensional   Lagrangian 
is still      
a   $s$-exact term, with a
topological gauge-function  that is not in 
 contradiction with the notion of ``holomorphicity". 
   Requirement 
of supersymmetry and explicit one-loop-order computations    
 provide  the explicit form of
$t_8$  \lerche\warner.
 We   suggest 
that  $t_8$ is a functional 
of $\Omega_4$. We find it appealing to believe   
  that
$\Omega_4$ is a reminder  of a propagating 
 four-form gauge field in ten dimensions, which has been  gauge-fixed 
 equal  to a
 background self-dual 
four-form in eight dimensions, using a ten-dimensional topological
symmetry.

Another way of  seeing the relevance of  higher-order corrections is to
compactify the eight-dimensional  theory down to a four-dimensional 
compact manifold,  say $K3$. If we integrate  out all massive modes,
this gives    a (non-local) four-dimensional gauge theory  for which
the infrared  description is given by an Abelian gauge theory with some
duality group $\Gamma$. The  possibility of defining the
eight-dimensional  theory on the product space $X \times K3$ in terms
of the  four-dimensional renormalization group might  lead us  to the
conclusion that there should be an analogous  Abelian  description in
eight dimensions. Quartic  terms contribute to the four-dimensional
prepotential upon compactification,  and we conclude  that the truly
eight-dimensional  features of the theory cannot be captured unless
these terms are  included in the discussion (plus corrections to all
orders in $\alpha'$) .
\foot{We would like to thank W. Lerche, S. Stieberger  and N. Warner for
pointing this out  to us and for 
important discussions and clarifications 
on the above questions pointed.}

The paper is organized as follows.
In Section 2,  
we   study in some detail  the  four-dimensional  situation. 
The formulation can be generalized to  that involving a $p$-form in $D$
dimensions and its   dual.
Our point of view is that
different gauge-fixings of the same topological symmetry, which involve
more fields, give the expression of the supersymmetric 
action either  in the non-Abelian (Donaldson--Witten) form  or in the 
Abelian (Seiberg--Witten) form, which   describe  the 
infrared behavior.

In
Section 3,  we   concentrate on a Yang--Mills field in the
eight-dimensional  case, and show how a dual formulation can be
obtained. In contrast with  the four-dimensional case,  
 the  
invariants of the
eight-dimensional case that 
  can  be constructed from the quadratic action should be 
expressible in terms of the classical ones.
However,  string corrections, which 
regularize the ultraviolet  behavior and 
enrich the infrared properties,  can open 
the gates to quantum effects.
Our results indicate the probable existence of a symplectic geometry
description  
analogous to that of the four-dimensional case. In particular, 
the gauge-fixing of the topological symmetry in the ghost sector
 leads us to
a relation between scalar components (defining the ``order parameters")
and their duals, which   generalizes that  of four-dimensional case. 
However, the  lack of a precise
renormalization group argument (which is replaced here by topological
arguments),
does not allow us to specify what   information is 
sufficient
 for
determining the      ``holomorphic prepotential", although 
a better understanding of string corrections could help solving this
issue. We also comment on the fact that from the 
eight-dimensional TQFT point of view, the
appearance of an extra monopole hypermultiplet in the 
low-energy theory of 
Seiberg--Witten is very natural, as first noticed in \bks.  
However, we do not discuss its  coupling to the genuine Yang--Mills TQFT 
in the  four-dimensional section, since the way to do it 
is obvious and  does not ad    new ingredient.

 Finally, in
Section 4, we indicate how our point of view also applies to 
two-dimensional duality, for a coupled scalar and Yang--Mills
supersymmetric theory. Obviously,  the dimensional reduction to
two dimensions of the    eight-dimensional theories in their various
formulations might be  independently interesting.

\newsec{Four dimensions}

\subsec{The fields and the ungauged-fixed action }

 \def\w{\wedge}
Let us consider a Yang--Mills field $A$ in four dimensions.
As explained in \kyoto \geneva, it is natural   to
associate  a two-form  gauge field $B_2$
with  $A$. This relies 
 on the unification  between   gauge fields and  ghosts, for the
fields as well as for their Batalin--Vilkoviski antifields.   Indeed, 
 in four dimensions, the ghost
expansion of  a two-form contains the
antifields
$A^{-1}_3$ and $c^{-2}_4$ of the gauge field
$A$ and of the Faddeev-Popov ghost $c$,  while that  of the Yang--Mills
field  $A$ contains the antifields $B^{-1}_2$,
$B^{-2}_3$ and $B^{-3}_4$ of the two-form $B_2$ and of its ghost and
ghost of ghost $\Psi^1_1$ and $\Phi^2_0$.

Our aim is to show    that the pairing of $A$ with a two-form $B_2$
gauge field provides  an   
 understanding of duality properties in four dimensions. Actually, 
 to write an Lagrangian, we must   
  introduce     another pair, $A_D$ and $B_{D2}$. This doubling
of degrees of freedom becomes clearer if $A$  is  replaced by a $p$-form
gauge field in an arbitrary dimension $D$, and if  we  generalize the
idea of ghost unification, as in \geneva. We   will go back to this
in Section  3, for the case of a Yang--Mills theory in eight dimensions, and in Section 4, which is devoted to the two-dimensional situation.

The one-form 
$A_D$ cannot be understood  as a Yang--Mills field since  
the  curvature $G_{A_D}$ of $A_D$ turns out to be its covariant
derivative 
with respect to $A$, that is, 
$G_{A_D}=D_A A_D$, with  
 $D_A =d +[A,.]$, while that of $A$ is  the Yang--Mills curvature 
$F_A=dA+ A\w A=   dA+\demi[A,A]$.

According to   \kyoto, the following expansion
determines the fields of the theory:
\def\A{{\tilde A}}
\def\F{{\tilde F}}
\def\B{{\tilde B}}
\def\U{{\tilde U}}
\def\V{{\tilde V}}
\def\X{{\tilde X}}
\def\Y{{\tilde Y}}
\def\W{{\tilde W}}
\def\Z{{\tilde Z}}

 \eqn\expA{\eqalign{\matrix{     &    
{  \A } &=&
{    c }  &+&
{    A}  &+&
{    B_{  2 }^{  -1 } }  &+&
{  \P_{  3 }^{  -2 }  }  &+&
{    \Phi_{  4 }^{  -3 }} 
\cr   
         &   
{  \B_{ 2} } &=&
{    c_4^{-2} }  &+&
{    A_{3}^{-1}}  &+&
{   B_{ 2  }^{    } }  &+&
{   \P_{  1 }^{ 1 }  }  &+&
{   \Phi_{  0}^{ 2 }}  }}}
 \eqn\expAD{\eqalign{\matrix{     &    
{  \A_D } &=&
{    c _D }  &+&
{    A_D }  &+&
{    B_{ D 2 }^{  -1 } }  &+&
{  \P_{ D 3 }^{  -2 }  }  &+&
{   \Phi_{ D 4 }^{  -3 }} 
\cr   
         &   
{  \B_{ {D2}} } &=&
{    c_{D 4}^{-2} }  &+&
{    A_{D3 }^{-1}}  &+&
{   B_{ {D2} }^{    } }  &+&
{   \P_{ D 1 }^{ 1 }  }  &+&
{   \Phi_{ D 0}^{ 2 }}  }}}
In this expansion, the forms with negative ghost number are antifields.
We follow the usual notation that the lower index is the ordinary form degree, which cannot exceed the value of the space dimension, while the upper index is the ghost number. 
The way we display the ghost  expansions in \expA\ and \expAD\
clearly  indicates the field-antifield pairings, as they were
sketched at the beginning of this section.

The symmetry of the theory is defined
by a BRST, that is, an
$s$-symmetry operation, where $s$ acts as a differential operator graded
by  the sum of ordinary form degree and ghost number, and
$s^2=0$. It turns out that
$s$ can  often be related to ordinary supersymmetry transformations by
the operation of twist.

This $s$ defines   classical  infinitesimal
transformations governed by two
one-form parameters $\rho$ and  $\rho_D$, which
    correspond to  the
ghosts  $\P_{   1 }^{ 1 }$ and $\P_{ D 1 }^{ 1 }$, and by two
 zero-form parameters $\e$ and $\e_D$,  which
    correspond to  the
ghosts  $ c$ and $c_D$. Here, $\e$ is the ordinary 
Yang--Mills transformation parameter, and  $c$ is the associated 
Faddeev--Popov ghost.

In \kyoto, a topological   $s$-symmetry for the fields in
 \expA\ and \expAD\
has been introduced, by means of   the master
equation of the field-antifield dependent Batalin--Vilkoviski Lagrangian:
 \def\w{\wedge}
\def\DA{{D_A}}

\def\DAT{{D_\A}}

\def\FA{{F_A}}
\def\FAT{{F_\A}}
 \def\L{{\cal L}}
 \def\I{{\cal I}}
\def\iF{{\cal F}}
\def\t{{\tau}}

\eqn\IUVp{\eqalign{\L  =   {\Tr}   \Big( 
\B _{D2}\w \B _2 +  
\B_2 \w  D_\A \A_D +
\B_{D2} \w   F_\A    
 \ \Big)\big|^0_4 . }}
We have
\eqn\IBV{\eqalign{s \int  \L=0,  \quad 
s\phi = {{\delta \int \L} \over {\delta\psi}} ,  \quad 
s\psi = {{\delta \int \L} \over {\delta\phi}}, }}
where, generically, $\psi$ is the antifield of $\phi$. Since \IUVp\ is a
Lagrangian of the first order, its equations of motion formally determine
  the BRST equations for all   fields and antifields.

A compact way of writing the action of $s$ on all
 fields and antifields, which leaves  the Lagrangian \IUVp\ invariant, is:
\eqn\brtsA{\eqalign{s\A&=-d\A-\A\w\A+\B\cr
s\B_2&=-\DAT\B _2\cr  
 s\A_D &=-d\A_D-[\A, \A_D]+\B_{D2}\cr
s\B_{D2}&=-D_\A \B_{D2}-[\A_D,\B_2]. }}
The $s$-variation of each one of the  fields and antifields is then 
obtained by a
further 
expansion in ghost number.  It give  in particular 
the topological transformations  $sA=\Psi^1_1+\ldots$ and 
$sA_D=\Psi^1_{D1}+\ldots$

Observing the way $\A$ and $\A_D$ transform, it is justified to examine 
if  we can add to the Lagrangian 
\IUVp\ a term 
$\iF =\iF(\B_2,\B_{D2})  $. Here, 
$\iF $ is a group scalar. 
By assumption, we chose it to be 
metric-independent.

The condition that we have an  $s$-invariant action, such that 
$s^2=0$, with a non-Abelian symmetry, implies that
$\iF $ is a function of $\B_2$ only:
$\iF =\iF(\B _2)  $. Moreover, 
$\iF $ must fulfil  the condition:
\eqn\cond{\eqalign{
\Big[\B_2, {{\delta \iF(\B_2)}\over {\delta \B_2
}} \Big] = 0 .
}}
This condition     holds in particular  when the gauge group is $SU(2)$.
Then, the new Lagrangian is:
\eqn\IUV{\eqalign{\L  = &  {\Tr}   \Big( 
 \B _2\w \B_{{D2}}\ +  
\B_2\w  D_\A \A_D  +
\B_{D2} \w   F_\A    
+  \iF(\B_2 )\cr
&+x \F_\A\w \F_\A   +
2y \F_\A\w  D_\A{\A_D}  
+z( D_\A{\A_D}\w  D_\A{\A_D} +F_A\w[\A_D,\A_D])
 \ \Big)\big|^0_4 .
 }}

For completeness, we have added to the Lagrangian    purely topological
terms, where $x$, $y$  and $z$ are 
  complex numbers. They can be adjusted in such a way    that,
eventually, the
$\theta$ parameter of the theory is  defined modulo $2\pi$.  We are
not interested in this issue, so we will set $x=y=z=0$ in the following.

The  modified  topological $s$-invariance for the Lagrangian \IUV\ is:
\eqn\brtsAF{\eqalign{s\A&=-F_\A+\B_2 
\cr
s\B_2&=-\DAT\B _2
\cr   s\A_D&=-D_\A\A_D +\B_{D2}
+{{\delta \iF(\B_2,\B_{D2})}\over { \delta \B_2 }}\cr
s\B_{D2} &=- D_\A \B _{D2} +[\B_2,\A_D]. }}

Obviously, the asymmetry between    $\B_2$ and $\B_{D2}$  comes from that between  
$\A$ and $\A_D$ in the non-Abelian case. Since  $\A_D$  rotates under
Yang--Mills symmetry   and   has also its own local gauge symmetry,
with ghost
$c_D$, it is not a standard Yang--Mills field.

Provided   \cond\ is verified, 
 there is no further restriction on the
$\B_2$ dependence of   $\iF$ to have $s^2=0$ on all
fields.  The 
usual physical requirement is that, after gauge-fixing, 
the action contains    
a
 Gaussian part that  is positive-definite.
This implies   ${\rm Im} \
\partial^2
\F > 0$.  The choice of a given $\iF$, as in \sw,  is 
based on dynamical  requirements,
which go
beyond    phase-space considerations.

If  the symmetry is purely    Abelian,  there
is a formal   symmetry between $A$ and $A_D$,
and     
$\cal F$ can  depend     on $\B_2 $ and $\B_{D2} $.
However, 
 throughout the paper, we   
consider that the commuting case is a limiting case of the non-Abelian
one, and we will consider  that  
$\iF$ is a function of $\B_2$ only.

\subsec{Non-Abelian case, with $\iF =\iF(\B_2 )$}

\def\aa{\Phi^2_0}

The Lagrangian \IUV\ can be considered as 
a rather sophisticated form of a classical  topological invariant.
For a non-vanishing $\iF$,  it actually
 depends on the fermions of the theory, that is  on the topological
ghosts and on the antifields. Indeed,  by Taylor expansion, we have: 
\eqn\miracle{\eqalign{   \iF(\B_2 )\big|^0_4=&\
 {\Tr}   \Big( 
\pa{\iF(\aa )}    c_4^{-2}
+\demi 
\pa ^2{\iF(\aa )}  (B_2\w B+2\Psi^1_1\w  A_3^{-1}) \cr & 
+\demi 
\pa^3{\iF(\aa )}  B_2\w  \Psi^1_1 \w  \Psi^1_1
+{{1}\over {24}}\pa^4 {\iF(\aa )}    \Psi^1_1 \w 
\Psi^1_1
\w  \Psi^1_1\w  \Psi^1_1
\Big).\cr 
}}
We use
the notation  
$ \pa={{\pa}\over {\pa \aa }}$. The apparent complexity of this
part of the  Lagrangian \IUV\
is quite analogous  to that of consistent anomalies, prior to the
understanding of descent equations. One sees that when   $\pa^2 \iF\neq
0$, there are interactions between     $\aa$ , $B_2$ and
$\Psi$, and, eventually,  $A$. This occurs  even in the commuting
case.

The cohomology of $s$, for its part with
ghost number $0$,  is empty. This is almost obvious if one looks at the
way $A$ and $A_D$ transform in \brtsA.
Thus, we expect that the
``classical'' Lagrangian \IUV\ is a closed-term, with a relation
 to a sort of
super Chern--Simons-term.

We have:
\eqn\miraclea{\eqalign{  (s-d) \iF(\B_2 )=0.}}
This can be seen from the way $\B_2$ transforms.

Furthermore, in the commuting limit, we have that 
$ \iF(\B_2 )\big|^0_4$ is by itself 
the sum of  
$d$-closed and $s$-exact terms. Indeed, in this case, we can
get easily the
above-mentioned   Chern--Simons formula. One  uses 
$\B_2= s \A-F_\A$, and $(s-d)^2=0$, and:  
 \eqn\miracleb{\eqalign{   \iF(\B_2 )=&\iF((s-d)\A )
\cr=&
\sum_{n=0}
^\infty
{{\iF^{(n)}(0)}\over {n!}} ((s-d)\A)^n\cr
=&
(s-d)\sum_{n=0}
^\infty
{{\iF^{(n)}(0)}\over {n!}} \A((s-d)\A)^{n-1} \cr
=&
(s-d)\Big( \A\sum_{n=0}
^\infty
{{\iF^{(n)}(0)}\over {n!}} \B^{n-1}_2\Big ) \cr
}}
Thus, we have  
 \eqn\miraclec {\eqalign{ 
\iF(\B_2 )\Big|^0_4=
s\Delta ^{-1}_4+d(...)
}}
and
\miracleb\ gives
\eqn\miracle{\eqalign{  \Delta ^{-1}_4 & = 
{\Tr}\Big( 
\pa{\iF(\aa )}    \Phi_4^{-3}
\cr &
+ {{1}\over {2}}
\pa ^2{\iF(\aa )}  \Big(\Psi ^{-2}_ {3} \w \Psi ^{1}_ {1}
+ B_2 \w B_2^{-1}   
+A \w A ^{-1}_ {3}  
+c c ^{-2}_ {4} \Big)
\cr & 
+ {{1}\over {6}}
\pa^3{\iF(\aa )} 
\Big (  \Psi^1_1\w  \Psi^1_1 \w B_2^{-1}
+2A\w B_2 \w \Psi^1_1
+c B_2\w B_2
+2c 
\Psi ^{1}_ {1}
\w
A ^{-1}_ {3}
\Big )
\cr & 
+{{1}\over {24}} \pa^4 {\iF(\aa )} 
\Big(  c    
\Psi^1_1
\w  \Psi^1_1
\w  \Psi^1_1
\w  \Psi^1_1
\Big)
\Big).\cr
}}
This expression is quite instructive, and shows the role of
antifields.    Let us indicate by anticipation that, after the
self-dual gauge-fixing that   provides,
by following the standard rules of  the Batalin--Vilkoviski formalism, 
an action that is suitable for path integration,
the 
antifields will  become functions of   propagating  antighosts.  
Actually, 
 the self-dual gauge determines  
  values for the
antifield of the following type:
\eqn\antifield{\eqalign{   
\Psi^{-2}_3  &=\  ^*(D _A\Phi^{-2}_{0})\cr
A^{-1}_3  &=\  ^*\big(   D _A\w ^*\chi^{ -1+}_{2}+
[\Phi^{-2}_{0},\Psi^{1}_{1}] +\lambda dc \big)
\cr 
\Phi^{-3}_4  &=\  ^*[\Phi^{-2}_0,\eta^{-1}_0]
\cr 
B^{-1}_2  &=\  c^{-2}_4=0.\cr
}}  
Here the constant  $\lambda$ is the usual gauge parameter
 for fixing the longitudinal
degrees of
freedom of $A$.

 Finally,  to directly  obtain    the action of $s$ on all
fields, it is convenient to decompose 
all fields in \IUV, and to use the interpretation 
of  the antifields as the
sources of the BRST transformation of fields as given by \IBV.  This gives:
\eqn\IUVaaa{\eqalign{\L  = \ &  {\Tr}   \Big( 
  B_2 \w  B_{D2}\ +  
 B _2\w  D_ A  A_D  +
 B_{D2} \w   F_ A 
   +  \t B_2\w B_2
\cr&
+A^{-1}_3\w(\Psi^1_{D1}+\t\Psi^1_{ 1} +D_A c_D -[c,A_D])
 +A^{-1}_{D 3}\w ( \Psi^1_{ 1}+D_A c)
\cr&
+c^{-2}_4(\Phi^2_{D0}-\pa \iF (\aa )-[c,c_D])
 +c^{-2}_{D 4}( \Phi^2_{ 0}-\demi[c,c])
\cr &
+B^{-1}_{ 2}\w(  D_\A \Psi^1_{D1}-[c,B_{D2}])
+B^{-1}_{D2}\w( D_\A \Psi^1_{ 1}-[c,B_{2}])
\cr  &
+\Psi^{-2}_{ 3}\w( D_\A \Phi^2_{D0}-[c,\Psi^1_{D1}])
+\Psi^{ -2}_{D 3}\w( D_\A \Phi^2_{ 0}-[c,\Psi^1_{ 1}])
\cr &
-\Phi^{-3}_{ 4}  [c,\Phi^2_{D0}]
-\Phi^{-3}_{D4}  [c,\Phi^2_{ 0}]
\cr & 
+\demi 
\pa^3{\iF(\aa )}  F_A\w  \Psi^1_1 \w  \Psi^1_1
+{{1}\over {24}}\pa^4 {\iF(\aa )}    \Psi^1_1 \w 
\Psi^1_1
\w  \Psi^1_1\w  \Psi^1_1
 \ \Big)  .
}}
We have defined the matrix:
\eqn\tauF{\eqalign{
\t &= {{\delta^2 \iF(\aa)}\over {\delta   \Phi^{ 2}_{ 0}
\delta\Phi^{ 2}_{ 0}
}}. }} 
 
\subsec{Restriction of fields in  the 
Cartan subalgebra}

Let us now consider the   possibility  of  using the 
topological symmetry to  do   a prior gauge-fixing, which   
reduces all fields   in the Cartan subalgebra of the gauge group.  
\foot{If the bundle is non-trivial, one must be careful in such a
``brutal" gauge-fixing". Indeed, it cannot 
 be decomposed as the sum of
line bundles, and it is not possible to globally choose the gauge
 that
reduces the theory to the Abelian one.  What one can do, instead, is to 
choose the Abelian
gauge on the complement to the finite number of points
(point-like  instantons); later, one needs to integrate over these
points. Interestingly, it seems that for some magic reason, the
gauge-fixing in the Cartan subalgebra, together with the introduction
of  the monopole hypermultiplet in our formalism, when we have a
non-vanishing ${\CF}$, is self-consistent, and the above remark can be
ignored. We wish to understand this better and will give some hint 
later from the eight-dimensional perspective.}

Once all fields are commuting,  there is 
a rotational symmetry   between $A$ and $A_D$, when $\iF=0$.
We  will
see that the  interpretation of Abelian duality is   the    freedom in
using different  gauge-fixings for  the  topological symmetry of both $A$ and $A_D$.

In the physical theory, the  moduli space of vacua is parametrized
by order parameters. These parameters are the  expectation values of
invariant polynomials of
$\Phi$; we denote the eigenvalues of the matrix $\Phi^2_0$ and 
 $\Phi^2_{0D}$ 
 by $a$ and $a_D$ respectively. 
The topological symmetry allows one to impose a relation  $c_D=0$, as a
gauge choice. A simple look at the way  $c_D$ transforms under $s$,
 ($s c_D$ is given by
the term in factor of $c_4$ in  
\IUVaaa),
indicates that  the 
 BRST invariance  implies a relation between the 
ghosts of ghosts (the scalars in   supersymmetry language)
\eqn\meand{\eqalign{
 a_D= {{\partial {\cal F}}\over \pa {a}}.
}}
The choice of another combination of $c$ and $c_D$, which one 
would gauge-fix to zero 
instead of $c_D$, is also possible.
 
\subsec{The topological invariance and the gauge-fixing freedom}

 To trigger intuition, consider   the   classical Lagrangian
$\L_{cl}$, obtained by 
 putting  all ghosts and antifields equal to zero in \IUV\    
and assuming that $\tau$ is constant: 

\eqn\Icl{\eqalign{\L _{cl}=   {\Tr}   \Big(  
  B _2\w   B_{{D2}}\ +  
 B_2 \w  D_A A_D    + 
 B_{D2} \w  F_A 
+    \demi B_2 \w\t(a)  B_2
 \Big)  
 }}
This  Lagrangian,   after elimination of $B_2$ and $B_{D2}$,  is $d$-closed.
It  can be called a  classical  topological term, and it  needs a
BRST-invariant  gauge-fixing.

 In the case of interest, that is  when $\tau$ is not a constant, 
  \IUV\   is  the sum of a $d$-closed {\it and }
an  $s$-exact term.  The latter cannot 
be interpreted as part   of  a gauge-fixing
term, since it   only involves  the  fields and  antifields of the
geometrical part of the BRST symmetry, prior  to  the introduction of
the cohomologically trivial  antighost sector.
It follows that  \IUV\ extends in a supersymmetric way 
the definition of a topological term.  We are   in a context that  is
slightly more general than that of  the standard Yang--Mills TQFT \bs,
which relies on the BRST invariant  quantization of an   ordinary topological  term.   However,
  the invariance of the Lagrangian \IUV\  still corresponds to 
the  invariances under
arbitrary shifts of $A$ and
$A_D$, as can be seen from the BRST equations \brtsAF. Thus as far
as the
gauge-fixing is concerned, we can extend the strategy as in \bs,
which can be applied whether $\tau$ is field-dependent or not.
For a  consistent gauge-fixing, one must use    the  complete action,
with its full  antifield dependence,  which    
  encodes all information about the BRST symmetry.

As already indicated, the novelty is that   \IUV\  can be
gauge-fixed in   different ways, which give 
various supersymmetric formulations that  can be  related by duality transformations.
  This property 
relies on the
following characteristics:

-- The symmetry on the gauge fields $A$ and $A_D$ is of the topological
type, as shown in \brtsAF;  the dependence in the two-forms $B_2$ and
$B_{D2}$ is purely algebraic, so these fields can be eliminated from the
action.

-- There are several ways to gauge-fix the symmetries on $A$ and $A_D$,
which leave
actions of the same type, but with different values of the coupling
constants.

All these properties hold at the Abelian as well as at the non-Abelian
levels and give different ways of expressing the theory. We can call
this property duality {\it covariance}; however, 
 duality {\it invariance}  only holds if 
 $A$ and $A_D$ are  valued  in the Cartan
subalgebra of the gauge group, a condition that can be realized by 
using  only   part of the freedom of the topological symmetry. In the
following, we will nevertheless write formula that accommodate the
non-Abelian case.

Let us make a technical comment. There are two ways of enforcing  
gauge conditions:
either one replaces the antifields in the Lagrangian \IUV\
by   relevant expressions     of
the type
$ \psi= {{\delta  Z^{-1}}\over \delta \phi}$, where 
the    chosen  gauge functions determine
$Z^{-1}$, according to the standard Batalin--Vilkoviski construction; or 
one adds  an
$s$-exact term to the   Lagrangian  obtained by setting  all
antifields equal to zero. The latter way is consistent because, for the type of symmetry that we consider,  the
antifield dependence is a linear one. Furthermore it is faster, since it
avoids the step of introducing the antifields of antighosts.  The
$s$-exact term also depends on the chosen  gauge functions. Both
procedures  are actually  equivalent and  necessitate    the
definition of a given set of  antighosts and of their Lagrange
multipliers. The antighost sector, which is BRST cohomologically
trivial, must be  adapted to the gauge choice.  As shown in
\bs,    freedom in the possibility of this sector of the theory   
makes the richness of TQFTs. We will  elaborate on this in the next
section, since this is precisely the possibility of having   different
classes of gauge-fixings  for different combinations of $\A$ and $\A_D$
which, will turn out to be  the key to obtain mirror-, or
duality-, related formulations.

It is worth noting at this point   that the
Seiberg--Witten holomorphicity properties  of the functional $\cal F$ 
are  understood in the BRST language as follows:    general  arguments
indicate that  the BRST cohomology cannot depend on antighosts. Thus it
is  expected  that any dependence of $\cal F$ 
  on the antighosts $\Phi^{-2}_0$ and $\Phi^{-2}_{D0}$, which we
shall shortly define, could be absorbed in an irrelevant counterterm. 
Since $\Phi^{-2}_0$ and $\Phi^{-2}_{D0}$ can  be interpreted as the complex
conjugates of  $\Phi^{ 2}_0$ and $\Phi^{ 2}_{D0}$ in the language of untwisted
supersymmetry, the property that $\iF$ can only depend on the latter
fields  finally gives the holomorphicity property.

\subsec {The gauge-fixing process}

The gauge-fixing of all components in $A$ and $A_D$ can be
done in various
ways, since we have   as many parameters in the symmetry as there are
modes in the gauge field (up to global excitations), see \bs. 
Let us now define the antighosts and Lagrange multipliers that are
needed for this purpose in  the $A$,
$B_2$ sector. The mirror equations for   the 
$A_D$, $B_{D2}$ sector   are obvious to deduce, and we skip   writing
them out.

As for the gauge-fixing of longitudinal modes in $A$, we have the
Faddeev--Popov antighost
$\bar c$, and its Lagrange multiplier $b$, and
$s\bar c=b$.  In the topological sector, we have antighosts and
Lagrange multipliers that complete the   ghost  spectrum
as follows: 
 \eqn\spectre{\eqalign{\matrix{         
  \  &
\ &
B_2 &
\ &
\  \cr
\  &
\Psi^{1}_{1} &
  &
\Psi^{-1  }_{1} &
\  \cr
  \Phi^{ 2  }_{0}  &
\ &
 \Phi^{0  }_{0}  &
\ &
 \Phi^{ -2  }_{0}   \cr
}}}
and 
\eqn\spectrel{\eqalign{\matrix{         
  \ &
H^{0  }_{1}&
\
\cr
\eta^{ 1  }_{0}   
  &
\ &
\eta^{-1  }_{0}    \cr 
}}}

The way $s$ acts on the antighost sector is BRST-trivial. It is:
$s \Psi^{-1  }_{1}= H^{0  }_{1}-[c,\Psi^{-1  }_{1}]$,
 $s  H^{0  }_{1}=
-[\Phi^{ 2 }_{0},\Psi^{-1  }_{1} ]$,
$s   \Phi^{-2  }_{0}= \eta^{-1  }_{0} -[c,\Phi^{-2  }_{0}]$,
 $s     \eta^{-1  }_{0}=
-[\Phi^{ 2 }_{0},\Phi^{-2  }_{0} ]$,
$s  \Phi^{0  }_{0}= \eta^{ 1  }_{0} -[c,\Phi^{0  }_{0}] $ 
and
 $s    \eta^{ 1  }_{0}=
-[\Phi^{ 2 }_{0},\Phi^{0  }_{0} ] $. 

A detailed treatment would imply the introduction of antifields for
the antighosts, (e.g.   $    \Psi^{ 0  }_{3}$ for $   \Psi^{-1  }_{1}$,
and so on),
  such that  terms can be added to \IUVaaa, which determine 
the    BRST
equations of the antighost sector from a master equation. It is
needless to
write   here  such terms,  which are of the 
type ${\Tr}  \Psi^{ 0  }_{3} \w (H^{0  }_{1} 
-[c,\Psi^{-1  }_{1}]  )  $.

\subsec{``Trivial" gauge-fixing} 

  \def\ww{{(w)}}

The first possibility of gauge-fixing that uses the 
topological freedom is the ``trivial'' one: we
 gauge-fix to zero   components of  $A$ and/or $A_D$, (which 
we will denote  by $A_{\ww}$ for notational simplicity).

This  gauge-fixing is obtained 
  by   the following $s$-exact Lagrangian,
which gives  algebraic terms that automatically 
gauge-fix to zero relevant ghost combinations for    the
$A_{\ww}$-sector: 
\eqn\trivial{\eqalign{       
 s {\Tr} \big( \Psi^{-1\mu  }_{{\ww} }  A_{{\ww}\m}      +\bar c
_{\ww}
\Phi^{0  }_{{\ww}0} &
+\Phi^{-2  }_{0} c_{\ww} \big)\ 
= \Tr \big (
H^{\mu}_{\ww}   A_{{\ww}\m}   +b _{\ww}\Phi^{0  }_{{\ww}0} 
 \cr
&+\eta^{-1  }_{{\ww}0}c _{\ww} 
+\Psi^{-1  }_{{\ww}\mu}  sA_{{\ww}\m}
  +\eta^{ 1  }_{{\ww}0}\bar c_{\ww}   
+\Phi^{-2  }_{{\ww}0}s c_{\ww} \ \big )\cr
}}

 We will use the gauge $A_{\ww}=A_D=0$ and $c_{\ww}=c_D=0$.
 Since, from \IUVaaa, $sA_D\big|_{c=0}=   
\Psi^1_{D1}+     \tau
\Psi^1_{ 1}$, and 
$s{c_D} \big|_{c_D=0}=\Phi^2_{D0}-\pa \iF (\Phi^2_{ 0})$, 
the integration  over the antighosts
$\Psi^{-1  }_{\mu}$ and $\Phi^{-2  }_{0}$ in \trivial\ gives
the  standard relations  $ \Phi^2_{D0}=\pa \iF (\Phi^2_{ 0})$
and
$  
\Psi^1_{D1} =     - \tau
\Psi^1_{ 1}   $.  These equations  shows  that, because of 
 the trivial gauge-fixing,  the BRST symmetry relates
the ghosts of dual
formulations  through    Legendre-transform type formula.

The gauge-fixing in the cartan subalgebra of $A$ and $A_D$ that we already discussed in Section 2.3 is of the ``trivial" type.

\subsec{Self-dual gauge-fixing}

After having ``trivially'' eliminated part of the fields, (for instance $A_D$),  we   can     use the
remaining freedom      to impose  a  self-dual equation 
on the curvature
of the remaining ones, that we will denote as $A_e$.  As explained   in \bs, 
the self-dual gauge-fixing  must be 
completed by an ordinary
gauge choice for the longitudinal modes.
Altogether, this gives four conditions on
$A_e$ and  exhausts all possible gauge freedom of the system.

We must now introduce, in the context of BRST invariance, 
the   gauge functions 
$F_{A_e}^+=(d{A_e}+{A_e}{A_e})^+$ and $\pa^\m A_{e\m}$.
For this purpose, we must  define our notation for the self-dual and 
anti-self-dual parts of a two-form: \eqn\duality{\eqalign{       
 B_2^\pm= \demi(B_2 \pm^*B_2)
}}
with (in Euclidian $4D$-space)
\def\s{\sigma}
$$
{^*B} _{\m\n}= {{\sqrt g \e}\over{2}} \e ^{\m\n\r\s}B_{\r\s},
$$
and thus, $B_2^+\w B_2^-=0$, and 
\eqn\dua{\eqalign{\matrix{       
 B_2\w {^*B}_2&=  d^4x \ \sqrt g \ { B}_ {\m\n}{ B}^ {\m\n}&=&
 B_2^+\w B_2^+
+
 B_2^-\w B_2^-
\cr
 B_2\w { B}_2&=   d^4x  \ \sqrt g \  { B}_ {\m\n}\e
^{\m\n\r\s}B_{\r\s}&=&
 B_2^+\w B_2^+
-
 B_2^-\w B_2^-\cr
}}}
The squared gauge function  $(d{A_e}+{A_e}{A_e})^+$ provides 
a Yang--Mills type Lagrangian plus a topological term, since ${^*}{^*}=1$,
and 
\eqn\id{\eqalign{\demi F_{A_e}^+\w F_{A_e}^+= F_{A_e} \w ^*F_{A_e} 
+
F_{A_e}\w F_{A_e}   .
}}

 To    enforce the  self-duality gauge function, we only need 
three antighosts and three  Lagrange multipliers.  We thus  need some
redefinitions of the    degrees of
freedom in the antighost sector.  (This eventually leads to the
possibility  of
twist.) We decompose:
\eqn\dua{\eqalign{ \Psi^{-1}_1 &\to (\chi^{-1+}_2, \Psi^{-1})\cr
H^{-1}_ 1 &\to (H^+_2, H),\cr
}}
where $\chi^{-1+}_2$ and $H^+_2$ are self-dual two-forms, and 
$\Psi^{-1}$ and $H$ are zero-forms. (From now on we skip the subindex
$e$ for the sake of notational clarity). Changing   variables as in
\dua\ 
amounts  to decomposing   a spinorial
tensor of rank $2$  into its  trace and its traceless parts.

One can eliminate the useless antighosts  $\Psi^{-1}$ and $\Phi^{0}_0$ 
by means of the
$s$-exact Lagrangian
\eqn\trivial{\eqalign{       
 s \Tr \big( \Psi^{-1  }   \Phi^{0  }_{0} \big)
=\Tr \big(
H  \Phi^{0  }_{0} +
 \eta^{ 1  }\Psi^{-1}\big ).
}}

The remaining part of \spectre\ is: 
  \eqn\spectred{\eqalign{\matrix{         
  \  &
\ &
B_2&
\ &
\  \cr
\  &
\Psi^{1}_{1 } &
 \  &
\chi^{-1+  }_{2} &
\  \cr
  \Phi^{ 2  }_{0}  &
\ &
 \   &
\ &
 \Phi^{- 2  }_{0}   \cr
}}}

 \eqn\spectreddd{\eqalign{\matrix{ \ &
H^+_2&
\  \cr
\  & \ &
\eta^{-1}_{0 }   }}}

 As explained in \bs, the   gauge-fixing also implies that of the
 longitudinal components of
$\Psi^1_1$. 
The $s$-exact term that enforces this condition is:
\eqn\gf{\eqalign{ s\ {\Tr}   \Big( \t(\Phi^{ 2}_0)
\chi^{-1}_{\m\n}{\big (
F_A^{+\m\n}+\demi H^{+\m\n} \big )
+\Phi^{-2}_0\big( D_A ^\m  \Psi^1_\m +[\Phi^{ 2}_0,\eta^{-1}_0]\big)
 \Big)   .
 }}
}
Finally, the gauge-fixing part of the longitudinal degrees of freedom in
$A$
 must be done by using the ordinary Faddeev--Popov ghost $\bar c$ and its
Lagrange multiplier.

\subsec{ The elimination of $B_2$ and $B_{D2}$}

The classical two-forms $B_2$ and $B_{D2}$ can be eliminated from the  
Lagrangian \IUVaaa\ by Gaussian integrations, for any given choice of the
function $\iF$. 
If we   put  all  antifields to zero, this gives a   term that  is a
quadratic form in the curvatures of 
$A$ and $A_D$, plus terms depending on the fermions $\Psi^1_1$ and the
derivatives of $\tau$. In what follows we do not consider the
dependence on the  fermions $\Psi^1_1$, although they are important  as
explained in \lns, in order to completely demonstrate the mapping
between the dual formulations.  We refer the reader to \wittendual  \mw  \lns\ for more  complete elements of the proof than the ones displayed below, which   include  the verification that the 
fermionic part of the action, obtained in expanding the $s$-exact terms
which enforce the self-duality gauge-fixing, obey the equivalence
between the actions.

The logic is as  follows. We   first perform a  gauge-fixing of all
fields in   the Cartan-subalgebra (the symbol $\Tr$ now means the trace in this space).   After this, there are two possibilities:
either  we impose the trivial gauge conditions $A_D=0$ and we recover  the standard topological term for $A$, by the elimination of the auxiliary fields $B_2$ and $B_{D2}$; or we first integrate out $A$, (this is the point where it is important to have done a prior  trivial gauge-fixing of all fields in the  Cartan-subalgebra). This implies that $B_{D2}=d\Lambda$, for some one-form $\Lambda$, and we can set $\Lambda=0$ by using the topological gauge freedom on $A$, which is not used  at the level of the equation of motion $dB_{D2}=0$.
Doing so,  $A $ and all its ghosts disappear. This procedure gives
Lagrangians that are  identical
up to the change
$\tau \to -1/\tau$ and field redefinitions.

Let us be a little bit more precise. Starting from  the  
Lagrangian \IUVaaa, with the above-mentioned restrictions,  the integration over $B_2$ gives:
\eqn\IclF{\eqalign{\L  \sim    & {\Tr}   \Big(  
 -\demi (B_{D2}   + d A_D  )\w   \t^{-1} (  B_{D2}+d A_D  )
+ B_{D2}\w F_A
\cr&
= {\Tr}   \Big(
-\demi   B_{D2}       \w   \t^{-1}    B_{D2} 
+ B_{D2}\w( F_A -\t^{-1}   d A_D )
 -\demi dA_D\w \t^{-1} d A_D
 \ \Big)   
 }}
If we now use the topological freedom to first gauge-fix $A_D=0$, we
obtain by   Gaussian integration: 
\eqn\IclF{\eqalign{\L   &\sim  
{\Tr}   \Big(  
 -\demi  B_{D2}    \w   \t^{-1}    B_{D2} 
+ B_{D2}\w F_A \Big )
\cr&
\sim  
{\Tr}   \Big(  
\demi F_A     \w   \t      F_A  
\Big )
}} 
This   is the standard topological term
which leads us directly to the    $\t$-dependent Yang--Mills
TQFT, by self-dual gauge fixing, as in \bs.

However, duality emerges because   we can use in a different and more refined way  
the topological freedom on
$A$  and $A_D$, by  gauge-fixing $B_{D2}$ to zero     after integrating out $A$.
Starting again from \IUVaaa, the  $A$-integration gives, together with summing over its fluxes:
\eqn\IclFaie{\eqalign{\L  \sim    & {\Tr}   \Big(  
   B_{2}\w(d\Lambda+dA_D)    
+  \demi B_{ 2}\w \t B_{ 2}
  \Big),   
 }}
where there is a functional integration over $\Lambda$. The latter field can be gauge-fixed to zero, using the topological freedom on $A$ which has yet not been used, and which gives
$s\Lambda=\Psi^1_1$.  Once $\Lambda=0$, the integration over $B_{D2}$ gives:
\eqn\IclFaie{\eqalign{\L  \sim    & {\Tr}      
 \Big (-\demi dA_D\w \t^{-1} dA_D\Big )
 }}
This Lagrangian, after self-dual gauge fixing of $A_D$, determines the dual theory, with the symmetry $\t\to-1/t$.

\newsec {Eight dimensions}


In this section, we will give the description of the 
 eight-dimensional Yang--Mills TQFT 
by extending what  we have done  in the
four-dimensional case. This TQFT  has been introduced in \bks,
 in a non-Abelian
formulation which is analogous to that of the
four-dimensional  Donaldson--Witten TQFT. Here, we will investigate
how to introduce the idea of duality in this theory.

In eight dimensions, the   dual of a one-form is a five-form.
If one considers the generalization of the 
 four-dimensional case, we        
have a pair  $A$,$B_6$, instead of $A$,$B_{D2}$, and   a pair
$A_5$,$B_2$, instead of $A_D$,$B_{6}$, with the following
ghost expansions for the fields and antifields that are   adapted to
the    eight-dimensional case:
  \eqn\expAA{\eqalign{          
  \A  & =
     c   +
     A  +
 B_{  2 }^{  -1 }    +
\P_{  3 }^{  -2 }     +
\Phi_{  4 }^{  -3 } + 
\Phi_{  5 }^{  -4 } +
\Phi_{  6 }^{  -5 } +
\Phi_{  7 }^{  -6 } +
\Phi_{  8 }^{  -7 }  
\cr   
     \B_{6 }&   =
     c_8^{-2}   +
     A_{7}^{-1}   +
    B_{ 6  }^{    }    +
    \P_{  5 }^{ 1 }     +
    \Phi_{  4}^{ 2 }    +
    \Phi_{  3}^{ 3}    +
    \Phi_{  2}^{ 4 }    +
    \Phi_{  1}^{ 5 }   +
    \Phi_{  0}^{ 6 } 
}}
 \eqn\expAAD{\eqalign{          
  \A_{5}  & =
    A_{0}^5 +
    A_{1}^4 +
    A_{2}^3  +
   A_{3}^2  +
   A_{4}^1  +
   A_{5}  +
\P_{ 6 }^{  -1 }     +
\Phi_{7 }^{  -2 } + 
\Phi_{ 8 }^{  -3 }  
\cr   
     \B_{2 }&   =
     A_{ 8}^{-6}   +
     A_{ 7}^{-5}   +
    A_{  6  }^{ - 4  }    +
    A_{   5 }^{ -3 }     +
    A_{   4}^{- 2 }    +
   A_{  3}^{-1 }    +
    B_{  2}^{   }    +
    B_{   1}^{ 1 }   +
    B_{   0}^{ 2 } 
}}
The Batalin--Vilkoviski Lagrangian  which generalizes \IUVp\ is:
\def\DA{{D_A}}

\def\DAT{{D_\A}}

\def\FA{{F_A}}
\def\FAT{{F_\A}}
 \def\L{{\cal L}}
 \def\I{{\cal I}}
\def\iF{{\cal F}}
\def\t{{\tau}}

\eqn\IUV{\eqalign{\L  =   {\Tr}   \Big( 
\B _2\w \B_{6}\ +  
\B_2 \w D_\A  \A_5   +
\B_6 \w   F_\A    
 \ \Big)\big|^0_8 . }}

 As in the four-dimensional case, this Lagrangian is invariant under a
topological symmetry
for  the Yang--Mills field $A$ and the five-form gauge field $A_5$, qiven by eq. \IBV. Its
classical part is equivalent to the topological density
$ {\Tr}   \Big(  D_ A A_5   \w  F_A  
 \ \Big)
$. Moreover, it is clear that we can introduce a 
``prepotential" $\iF(\B_2)$, add it to the Lagrangian and distort the BRST
symmetry as we did  in four dimensions. 
 
\def\O{\Omega}
 
It     seems, however, that  duality  cannot hold in eight dimensions
because of the obvious off-shell asymmetry between   the
Yang--Mills one-form $A$  and the   five-form $A_5$. Actually, this 
 difficulty can be circumvented,   because of the special properties of
eight-manifolds, which have been already used in
\bks\Acharya to make an eight-dimensional  Yang--Mills TQFT. Before
introducing  a prepotential, we must explain this.

 Assuming that the eight-manifold has $Spin(7)$ holonomy, there is a
canonical self-dual closed
 four-form
$\Omega_4(x)$ which is covariantly constant. It can be locally written
as follows: we choose a local vielbein such  that the metric is $\sum
e_i
\otimes
\e_i$,
where the $e_i$'s are one-forms and $i=1,...,8$; then
$\Omega_4=T^{ijkl}e_i \wedge \e_j \wedge e_k \wedge e_l$. This form is
invariant under the  rotations of $e_i$, which build  the
subgroup  $Spin(7)$
 of $SO(8)$.  The tensor  $T$ is self-dual, 
$T_{\m\n\r\s}=\e_{\m\n\r\s\a\beta\gamma\d}T^{\a\beta\gamma\d}$, and it
can be
written in terms of the  octonionic structure constants that define the
$G_2$-structure
 of  seven-dimensional   manifolds. Such    manifolds have a 
unique covariantly constant
three-form $\phi_3$, with its dual $\Omega_4$,
 which can be locally written
as    
$\phi_3=c^{abc} e_a \wedge
e_b \wedge e_c$, $a,b,c=1,...,7$, where
 the $c^{abc}$ are octonionic structure constants.
$\phi_3$ is invariant under the subgroup $G_2$  of
$SO(7)$, and $\Omega_4 = e_8 \wedge
\phi - *\phi$.

The invariant four-form
$\Omega_4(x)$ can be used to decompose  any given two-form $B_2$  into
two $Spin(7)$-irreducible components $B_2^\pm$, according to
$28=21\oplus 7$. $B^+_2$ and   $B^-_2$ can be called   self-dual 
   and antiself-dual respectively. The generalization
of the four-dimensional decomposition \duality\ is:
\eqn\dualitye{\eqalign{       
 B_2^\pm= \demi(B_2 \pm^\dagger   B_2)
}}
with (in Euclidian space):
\eqn\dualityes{\eqalign{ 
{^\dagger B} _{\m\n}= {{\sqrt g  }\over{2}} \O _{\m\n\r\s}B^{\r\s},
}}

As in the four-dimensional case, $B_2^+\w B_2^-=0$, and:
\eqn\dua{\eqalign{\matrix{       
 B_2\w {^*B}_2&=  d^8x \ \sqrt g \ { B}_ {\m\n}{ B}^ {\m\n}&=&
 B_2^+\w B_2^+
+
 B_2^-\w B_2^-
\cr
 B_2\w {^ \dagger{ B}}_2&=   d^8x  \ \sqrt g \  { B}_ {\m\n}\O
^{\m\n\r\s}B_{\r\s}&=&
 B_2^+\w B_2^+
-
 B_2^-\w B_2^-\cr
}}}
Obviously, the symbols  $^*$ now denotes the ordinary Hodge  duality
operation in eight dimension.

Now comes the new features. Using  $\O_4$, one   can decompose the 
gauge field
$A_5$ according to:
\eqn\duao{\eqalign{       
 A_5 =   \O_4\w A_D+ (A_5 -\O\w A_D),
}}
where
the one-form gauge field $A_D$ is defined as: 
\eqn\duaoo{\eqalign{       
 A_D =  ^*(\O_4\w ^* A_5).
}}   

The idea is to first use  the topological symmetry to gauge-fix to zero
$(A_5 -\O\w A_D)$, that is, to  enforce the following gauge-fixing:
\eqn\duaoo{\eqalign{       
 A_5 =   \O_4\w   A_D.
}}

Actually, a simple counting of commuting and anticommuting degrees of
freedom in 
$ \A_5$ and $\B_6$, together with the properties that the ghosts in
$\B_6$ are topological ghosts for $  A_5$,
indicates that one can indeed perform the following gauge fixing
\eqn\duaoo{\eqalign{       
 \A_5 = &  \O_4\w  \A_D
\cr
\B_6 = &  \O_4\w  \B_{D2}
}}
  The gauge-fixing  of $\A_5$ down to a one-form $\A_D$, and 
of  $\B_6$ down to   $\B_{D2}$ means that
that,   some of the ghost  contained in $\A_5$ and  $\B_6$ are   
used, after having introduced all necessary antighosts. In the BRST
framework, they 
    disappear from the theory because of the  algebraic
equations of motion 
stemming from the BRST invariant way of
  enforcing \duaoo.

The degrees of freedom  which 
are contained in the expansions of
of $\A_5$ and $\B_6$, and   which  remain ungauged-fixed 
after the gauge-fixing of $A_5$ to $A_D$, can be effectively reorganized
in the following expansions:
\eqn\duaooD{\eqalign{  
 \A_5=\O_4\w \A_D  & =\O_4\w\big (
     c_D  +
     A_D  +
 B_{ D 2 }^{  -1 }    +
\P_{ D 3 }^{  -2 }     +
\Phi_{D  4 }^{  -3 } \big )   
}}
 \eqn\expAADDDD{\eqalign{          
  \B_6=\O_4\w\B_{D2 }&   =\O_4\w\big (
    c_{  D 4}^{- 2 }    +
   A_{ D 3}^{-1 }    +
    B_{ D 2}^{   }    +
    \Psi_{  D 1}^{ 1 }   +
    \Phi_{ D  0}^{ 2 } \big )}}
  Notice
that the classical forms $B_6$ and
$ B_{D 2}$ have the same number of degrees of freedom, but different
gauge symmetries.

 This gauge-fixing reduces the Lagrangian to:
\eqn\IUVD{\eqalign{\L  = &  {\Tr} \O_4\w  \Big( 
 \B_2 \w \B_{D2}\ +  
\B _2\w  D_\A \A_D  +
\B_{D2} \w   F_\A     +  \iF(\B_2 ) 
 \ \Big)\big|^0_8 
 }} 
This a multiplication by $\O_4$ of a Lagrangian which is very similar
to that we have studied  in four dimensions.
We have actually  incorporated   a term 
 $\O_4\w \iF(\B_2 )$. By assuming the  proportionality 
of this term 
to 
 the four-form $\O_4$, we  
eliminate the possibility that the Lagrangian contains topological
terms proportional to powers in $B_2$ higher than two. We will
 soon discuss the relevance of supressing this condition.  

For $\iF=0$, the classical part of  $\L$ is equivalent to the
topological term 
$\O\w {\Tr}F_A \w F_A $, that is,  the starting point of the TQFT in
\bks.  

For   $\iF\neq 0$, the situation is   analogous to that   in four
dimensions, but
   the trivial gauge-fixings  must   be combined to  self-dual  gauge-fixings 
  which are  specific to eight dimensions. The latter is realized by 
an $s$-exact term of the following type:
\def\Ae{A_e}
\eqn\gff{\eqalign{ \Big   \{ Q, {\Tr}   \Big( 
\t \chi^{-1+}_{\m\n}{\big (
F_{\Ae  }^{+\m\n}+\demi H^{+\m\n} \big )
+\Phi^{-2}_0\big( D_{\Ae } ^\m  \Psi^1_\m +[\Phi^{ 2}_0,\eta^{-1}_0]\big)
 \Big)   \Big   \} .
 }}
}

The symbol $^+$ is now defined as in \dualitye.
(In fact, as explained in \bks, there are $7$ independent degrees in
freedom
$\chi^{-1+}_{\m\n}$ and $H^{ +}_{\m\n}$, according to the 
$Spin(7)$-decomposition 
$\underline {28}=\underline {21}\oplus \underline {7}$.)  Using \dua,
one easily
sees, by repeating the steps that we detailed in the four-dimensional
case, that
the gauge-fixed action is $\int {{1}\over {g^2}
}(F_{\Ae }\w^*F_{\Ae }+\ldots$, where the
 value of $g^2$ depends on the
chosen combination of 
$A$ and $A_D$ for $\Ae$. The notation  $\ldots$ stands for terms that
make the action identical, up to twist, to that of the 
$D=8$ SSYM theory, that is  the dimensional reduction to eight
dimensions of 
$N=1$ $D=10$ SSYM theory.

As for the generalization of \meand, we   have exactly the same relation: 
\eqn\meandd{\eqalign{
  a_D= {{\partial {\cal F}}\over \pa {a}}
}}
This equation is consistent
 the fact that 
$\O_4\w\iF$ is  an eight-form.
Indeed,  within the BRST interpretation,    the mean values of the  
ghosts of ghosts,
$a_D= <\Phi^2_{D0}>$  and $a = <\Phi^2_{ 0}>$ are scalars with ghost
number two, that is, two-forms, and thus \meandd\ is dimensionally
meaningful.

We thus formally  find  that duality properties  can  hold   in the eight-dimensional
Yang--Mills TQFT. The parallel with the four-dimensional case is of
course striking. It is due the possible  proportionality of the Lagrangian to $\O_4$.   Abelian duality will be    obtained by   an
initial  gauge-fixing  of the fields in the Cartan subalgebra of the
gauge 
group. 

However, one can go further:
we have the possibility of adding to the Lagrangian \IUVD, which already includes the term 
$\O_4 \w\iF (\B_2)$, an  $SO(8)$-invariant
term ${\cal G}(\B_2)
\big|_8^0$, which
is not proportional to $\O_4$. This   
gives new
interactions to the theory, but no modification of the 
quadratic part of
the action. However, the Lagrangian  
depend on quartic interactions in $B_2$, and we now look how these terms can be handled in    procedures analogous to those that led us to dual Lagrangians in the four dimensional case. The Lagrangian is:
\eqn\IUVDe{\eqalign{\L  = &  {\Tr}\Big( \O_4\w  ( 
 \B_2 \w \B_{D2}\ +  
\B _2\w D_A\A_D  +
\B_{D2} \w   F_\A     +  \iF(\B_2 ) 
 \ )+ {\cal G}(\B_2) \Big) \big|^0_8 
 }} 

As before, we gauge-fix all fields in the Cartan subalgebra, and we still  do not include the fermion terms in our discussion, which are
 proportional to higher derivatives of $ {\cal F}$ and  {\cal G}. Then, as a first possibility,  
we can set   $A_D=0$.   After  elimination of $B_{D2}$, this gives  
a classical Lagrangian :
\eqn\IUVDee{\eqalign{\L  = &   {\Tr}\Big( \O_4\w     
   \iF(d A ) +  
 {\cal G}(d  A)  \Big)  
 }} 
 The latter can be gauge-fixed in the octonionic self-dual way as in     \bks.

As a second possibility, we first integrate over $A$. This gives, by summing over all fluxes, $B_6=d\Lambda_5= \O_4 \w d\Lambda $, and as a last step, we use the gauge freedom on $A_D$ to set 
 $\Lambda =0$. (The analysis about the way to use all ghosts is as in four dimensions). At this point, the Lagrangian is:  
\eqn\IUVDeee{\eqalign{\L  = &  {\Tr} \Big(\O_4\w  (
 B _2\w d  A_D +   
   \iF(B_2 ) \ ) +
 {\cal G} (B_2) \Big).   
 }} 
 If ${\cal G}=0$, the discussion is exactly    as in four dimensions, apart from the overall multiplication by $\O_4$.   If ${\cal G}\neq 0$, we cannot perform exactly the quartic  integration. However, one can treat these term as a perturbation, and replace, in a first approximation,  the argument of ${\cal G} $ by 
$B_2=\t^{-1}dA_D$. 

Now comes the physically interesting question. The  eight-dimensional 
theory is non-renormalizable and it is   infrared-free, as indicated by
power counting. From the other side,  the topological theory described
above is basically the same as the 
four-dimensional theory: its  Lagrangian is the product
of the 
$Spin(7)$-invariant 4-form $\O_4$ with a Lagrangian very similar to 
the four-dimensional one.   It is tempting to conclude, based on the
possibility of the  gauge-fixing procedure described above, that 
the  eight-dimensional
Abelian topological theory can give an  analogue  to the 
Seiberg--Witten ``infrared description". It should be kept in mind that
our eight-dimensional theory
is  not really topological from the beginning, since  only the metric
deformations that preserve the $Spin(7)$-structure are allowed. Thus
  arguments relating the ultraviolet  theory to the infrared
one    seem     difficult to use, although an ``Abelianization" can be 
achieved by a gauge-fixing which take into account the   
delicate questions of the treatment of pointlike instantons.  At present
we unfortunately lack  an  explicit derivation of the infrared 
finite-dimensional integral from the compactification of the instanton
moduli space even in four dimensions, see for example
\lns.  \foot {Note that,  although the ultraviolet TQFT
is not Lorentz-invariant in eight dimensions, since it explicitly
depends on  $\O_4$, its  untwisting  is possible  and
one recovers the ordinary supersymmetric  theory that is
perfectly Lorentz-invariant as  was shown in \bks.  If our
eight-dimensional   is a product of two four-manifolds, or if  it is 
  a
fibration over a four-dimensional base,  we can think about this theory
as a four-dimensional one, such that  the   integral 
$\int_{fibre}\Omega_4$ gives the coupling constant in four dimensions.
In this way,   we can recover the four-dimensional topological gauge
theory with all its properties.}

\def\sp{\kappa}
There is a question that we   ignored in our
four-dimensional discussion, and which concerns the coupling of
supersymmetric matter to the Abelian theory. Another    interest of
the   eight-dimensional Yang--Mills TQFT   is to   enlight    
this coupling, which
uses  a commuting spinor, and  
generalizes the genuine Yang--Mills TQFT by   a  
modification of the  four dimensional  
 self-duality gauge condition, which becomes the equality of the
self-dual part of the Yang--Mills curvature to   the commuting-spinor
current
\wittendual. As first noted in
\bks, 
the eight-dimensional Yang--Mills TQFT precisely gives, by dimensional reduction,
 a TQFT using   four-dimensional Seiberg--Witten equations as 
topological gauge functions. The argument is as follows. 
By dimensional reduction from eight to four dimensions, the
eight-dimensional  gauge field determines a four-dimensional gauge field, (made from the first four components $A_\mu$, $1\leq \mu\leq
4$), and  a four-dimensional bosonic complex Weyl spinor 
$\sp  $  (made from    the remaining four components   $5\leq\mu\leq
8$). Then,  the seven octonionic
 gauge functions, which determine the TQFT in eight dimensions,
 become four-dimensional ones:
\eqn\swe{\eqalign{
F^{a+}_{\mu\nu}&=  f^a_{bc}  {^{  +}\sp}^{b} \gamma_{\mu\nu} \sp^{c}
\cr
  \gamma^\mu D_\mu \sp ^a&=   0 .
}}
Here the indices $\mu,\nu$ run from $1$ 
to $4$, and $a, b,c$ are Lie algebra indices;
$f^a_{bc}$ are the Lie algebra structure constants.
These equations are non-Abelian, and differ from those
given in \wittendual\ in the commuting limit. Thus,   it 
is  
useful  to explain in more detail     how it could 
allow  us to
obtain the {\it Abelian} Seiberg--Witten  equations, for which $A$ is
just a
$U(1)$ gauge field, rather than \swe, where $A$ is  non-Abelian.  This
is indeed quite simple, at least formally. One must do   a 
first gauge-fixing
in the eight-dimensional theory 
which sets  equal to zero  the  components of 
$A^a_\mu$ with  $1\leq\mu\leq 4$ that    are not in the Cartan subalgebra,
  and,  for $\mu, \nu\geq 5$,  all combinations of
$A^a_\mu$, say,
${\cal  A}^a_\mu=\sum_b c^a_b A^{ b }_\mu$, that  are such that 
$f^a_{bc} {\cal A}^{b}_\mu {\cal A} ^{c}$
is also not  valued in the Cartan subalgebra.
This algebraic, or ``trivial'',
gauge-fixing in the Cartan subalgebra
also eliminates the corresponding ghosts in the TQFT.
Then, for the remaining fields, one 
 uses  the octonionic gauge function. This 
automatically gives   Abelian Seiberg--Witten equations.

It is best to give an example. Let us consider the $SU(2)$ case.
In the non-Abelian phase, both  $A$ and $\sp$ are   an $SU(2)$-triplet.
The gauge-fixing  sets $A^{(1)}=A^{(2)}=0$  and   
$\sp^{(3)}=\sp^{(1)}-\sp^{(2)}=0$. If we define $\sp=
\sp^{(1)}=\sp^{(2)}$, then the remaining topological gauge freedom on
$A^{(3)}$ and $\sp$ can be used, and     the octonionic gauge
function gives:
\eqn\swab{\eqalign{
F_{\mu\nu}^+  & =  {^{  +}\sp} \gamma^{\mu\nu} \sp 
\cr
\gamma^\mu D_\mu \sp & =  0,}}
where    $F=dA^{(3)}$ as a result of the first gauge-fixing on $A$.
This shows that the non-Abelian 
eight-dimensional 
Yang--Mills theory formally gives the 
four-dimensional TQFT
with  ``Abelian''
Seiberg--Witten gauge-functions, provided a preliminary gauge-fixing 
has been done   to restrict all  equations in the Cartan subalgebra of
the Yang--Mills  group.

Let us return to the eight-dimensional
case and see how the theory may depend on  
higher derivative couplings.  As an example, in the gauge-fixed
theory,
we may need a non-topological quartic term:
 
\eqn\ct{\eqalign{
t^{\m_1 \m_2\m_3\m_4\m_4\m_5\m_6\m_7\m_8}(\Phi^2_0)
F_  {\m_1 \m_2}
F_  {\m_3 \m_4}
F_  {\m_5 \m_6}
F_  {\m_7 \m_8} 
}}
 Such terms were computed
in 
\warner\ for string compactifications.
The appearance of the tensor $t$ is actually well known in string
theory
 \lerche.
It is made of the product of a certain trace of $\gamma$ matrices, and
it 
is related to the   four-form $\Omega_4$.

One can explain the relevance of terms as in \ct\ as follows: 
if one starts from the eight-dimensional action TQFT,    which
can be  twisted in an action with eight-dimensional supersymmetry, one
soon realizes   that non-topological        
quartic counterterm  such that \ct\ are  needed, for instance by
one-loop corrections.

From our view-point,
one should be able to   obtain 
such terms   by adding an $s$-exact counterterm,
which was absent at the tree-level 
\foot{An   example of the necessity of improving   $s$-exact terms  is when one 
renormalizes   the Yang--Mills 
theory in a non linear gauge: higher-order ghost interactions must be 
introduced by mean of a BRST-exact term, in order to compensate for the 
divergences in the four ghost vertex 
  occurring in such gauges. }. The
way to incorporate such terms is quite simple: 
we can  change  the definition of $F_{\Ae  }^{ \m\n}$ in the part \gff\
of the Lagrangian,  which enforces the self-duality gauge condition,  
as follows:  
\eqn\gffr{\eqalign{  F_{\Ae \m\n }^{  }=\pa_{[\m}A_{ e\n]} \to
 F_{\Ae r \m\n} { }
= 
F_{\Ae  \m\n}^{  }
+
t  _{ \m\n\a\b\c\d\r\s}(\Phi^2_0)
F_{\Ae  }^{ \a\b}
F_{\Ae  }^{ \c\d}
F_{\Ae  }^{ \r\s}
}}
Indeed,   squaring   the self-dual part   
$F_{\Ae r }^{ +\m\n}$ produces, in addition to the standard
 Yang--Mills Lagrangian, the term \ct.
 Additional fermionic terms are   generated: they are  very easy  to
find by expansion of  the new $s$-exact term.   There are of course 
 terms of degree $F^6$ that  are produced.
 They are of order $\alpha'^2$, 
thus corresponding string computations are required.
Actually, 
it is a very interesting question 
to find out whether  all string corrections that are needed 
to renormalize the theory  and that can be organized as 
a formal series in $\alpha'$ with higher powers of 
$F$, can be obtained by a replacement of $F_A$ in the self-duality
condition, which is analogous to \gffr. (It is possible that the Born--Infeld type Lagrangian can emerge in such a  process.) Here, we simply  note that   the duality considerations of this section suggest that,
  when both quadratic and quartic couplings are introduced
at the leading order, one has the relation: $\tau_D = - 1/\tau$, 
$t _{D8}\sim t_8 / \tau^4$. It would  be interesting to verify  if  the 
dual formulation  that we have introduced in this section fits in the context of  
  string theory,  together with these transformation rules (which 
unfortunately are not exact since  quartic interactions do not  allow
us to exactly  integrate out the auxiliary fields).

Finally, we recognize that having the four-form $\O_4$ is an essential
building block of the theory. It could be the eight-dimensional
projection of a self-dual propagating four-form of a ten-dimensional
theory. More precisely, we  may think of 
a ten-dimensional theory, whose action is of   
a Chern--Simons type,  function of   $\O_4$,  
 and of two two-forms gauge
fields $C_2$ and
$C'_2$, and has the following expression:
\eqn\ten{\eqalign{ 
\int_{10}
d\O_4 \w C_2\w dC'_2.
}}
This introduces the interesting case of theories that involve a 
five-form curvature $d\O_4+C_2\w dC'_2-C'_2\w  dC _2$, with a
non trivial dependence on consistent anomalies.

To summarize this section,  we have seen   that   
 the 
topological freedom for the large set of fields $A, A_5, B_2$ and $B_6$
allows a duality picture in eight dimensions, with
a clear relationship with the case of four dimensions.

\newsec{Two dimensions}

\def\e{\epsilon}

\def\demi{{1\over 2}}

\def\pa{\partial}

\def\a{\alpha}

\def\b{\beta}

\def\c{\gamma}

\def\m{\mu}

\def\n{\nu}

\def\r{\rho}

\def\L{L}

\def\X{X}

\def\V{V}

\def\P{\Psi}

\def\F{\Phi}

\def\d{\delta}

\def\Z{{\tilde Z}}

 \def\A{{\tilde A}}

\def\v{{\varphi}}

\def\B{{\tilde \v}}

\def\U{{\tilde U}}

\def\V{{\tilde V}}

\def\X{{\tilde X}}

\def\Y{{\tilde Y}}

\def\W{{\tilde W}}

\def\Z{{\tilde Z}}

    We can dimensionally reduce to two dimensions   the theories that we
have introduced in four and eight dimensions.
However, we can also
consider  our  duality machinery   directly in
two dimensions.

We  thus start with     a two-dimensional gauge field $A$. Its ghost
expansion is 
$
{  \A }=
{    c }  +
{    A}  +{    \v_{  2 }^{  -1 } }
$. Its dual is, formally, a form of degree $-1$,   $\Z_{-1}$, with  
 no classical content. Indeed, 
$\Z_{-1}$ can  only contains fields with negative ghost numbers,
$\Z_{-1}     = 
{ \F_{ 2 }^{ -3 } }+
{     \P_{  1 }^{ -2 }} +
{   W_{ 0 }^{-1    } }$. These fields  can 
  be identified as the   antifields   of a two-form  $W_2$,  of
its ghost $\P_{  1 }^{ 1 }$,  and its ghost  of ghost
  $\Phi_{  0 }^{ 2 }$. 
In turn, the existence of $ W_2$ implies that of its dual, which is a
scalar that we call $\v$. The existence of $\v$ could have been directly
inferred from that of 
$ \v_{  2 }^{  -1 }$ in $\A$. So, 
  introducing a two-dimensional gauge field $A$  
 leads us to   defining:
\eqn\expA{\eqalign{\matrix{         
{  \A }  =&
{    c }  +
{    A}  +{    \v_{  2 }^{  -1 } }  
\cr     
{  \B  } =&
{    c_2^{-2} } +
{    A_{1}^{-1}}  
+\v }}}
and
\eqn\expW{\eqalign{\matrix{ 
{  \Z_{-1} }   =&
{ \F_{ 2 }^{ -3 } }+
{     \P_{  1 }^{ -2 }} +
{   W_{ 0 }^{-1    } } \cr        
{  \W_2 }  =&
{   \F_0^2} +
{     \P_{  1 }^{  1 }}  +
{    W_{  2 }^{  } } .
 }}}

Now we also want   (string coordinates) scalars  that we call $X$; we
introduce their duals,  which we call    $Y$ (we could have denoted
$Y=X_D$). The additional one-forms, which   play a role analogous to
that  of the two-forms $B_2$ and $B_{D2}$ in four dimensions   and 
contain in their ghost expansion  the antifields of $X$ and $Y$,  are
called
$U_1$ and
$V_1$. So, in the string coordinate sector,  we can  define two sets of
``dual'' coordinates:

\eqn\expU{\eqalign{\matrix{         
{  \X }  &=&
{   X } 
      +     U_{   1 }^{ - 1}    + 
  U_{  2  }^{ -2 }    
\cr   
  \U_{1}     &=&
 X_{  2 }^{  -1 }  +
      U_{ 1 }^{   } 
+U_{ 0 }^{ 1 } 
\cr}
 }}

\eqn\expU{\eqalign{\matrix{         
{  \Y }  &=&
{   Y  }+
            V_{   1 }^{ - 1}    + 
  V_{  2  }^{ -2 }    
\cr   
  \V_{1}     &=&
Y_{  2 }^{  -1 }  +
      V_{ 1 }^{   } 
+V_{ 0 }^{ 1 }. 
\cr}
 }} 

We can  now introduce a TQFT Lagrangian density as the 
following two-form
with ghost number zero: 
\def\w{\wedge}\def\DA{{D_A}}
\def\DAT{{D_\A}}
\def\FA{{F_A}}
\def\FAT{{F_\A}}
 \def\L{{\cal L}}
 \def\I{{\cal I}}
\eqn\IUV{\eqalign{\L_2 =   {\Tr} \ \Big(\ &
\U_{1} \w \V_{1}\ +\ 
\U_{1} \w  \DAT \X  \ +\ 
\V_{1}  \w \DAT \Y_{ }
 \cr &
+\B_{ } \w \W_2 
+  \B_{ } \w \FAT
+ \W_2 \w 
(\DAT \Z_{ -1} +[\X ,\Y_{ }]
\ )\ \Big)\ 
\Big| ^0_2\ . }}
This Lagrangian    is of  first order
and metric-independent; its purely classical  part  $\L_{cl,2}$  is:
\eqn\IUVcl{\eqalign{\L_{cl,2} =   {\Tr}  \Big(\ &
  U_1\w V_1
+U_1\w \DA   X{} 
+V_1  \w \DA   Y_{}
+ \v (W_2+\FA)
 +W_2 [ X{}, Y_{}]  \Big) . }}
If we  eliminate  
$ X_{ 1}$,
$ Y_{{ 1}}$ 
 and 
$\v$
by their algebraic equations of motion $Y=\DA X$, $X=\DA Y$ and $W=\FA $,
we obtain:

\eqn\IUVcl{\eqalign{\L_{cl,2} \sim    {\Tr} \ \Big(  
  \DA   X{}   \w \DA   Y_{}
 +\FA \w [ X{}, Y_{}]  \Big) . }}
We see that \IUVcl\ is   closed, contrary to
\IUV.  The integral over the two-dimensional space
of the density 
\IUVcl\ can be considered as a topological term. This indicates that  
we are  dealing with  the Lagrangian of a TQFT, and we can repeat the
same manipulations as led us to duality in four dimensions.

The   invariant Batalin--Vilkoviski-type action is:
\eqn\IBV{\eqalign{\I_2=\int _2 \L_2 . }}
which satisfies the master equation \IBV.

Thus, \IUV\ is invariant under the topological   BRST symmetry:   
\eqn\brtsA{\eqalign{s\A&=-\FAT+\W_2\cr
s\W_2&=-\DAT\W_2 . }}
\eqn\brtsU{\eqalign{s\X{{}}&=-\DAT \X_{}+\V_{{ 1}}\cr
s\V_{{ 1}}&=-\DAT\V_{{ 1}}. }}
\eqn\brtsV{\eqalign{
s\Y_{}&=-\DAT \Y_{}+\U_{ 1}\cr
s\U_{ 1}&=-\DAT\U_{ 1}. }}
\eqn\brtsB{\eqalign{ 
s\Z_{  -1}&=-\DAT\Z_{  -1}+[\X_{ }, \Y_{}] +\B_{{  }}
\cr
s\B_{{  }}&=-\DAT \B_{{  }}
+[\U_{ 1},\X{}]
+[\V_{{ 1}}, \Y_{}]
+[\W_{2}, \Z_{  -1}]
. }} 

The situation is very reminiscent of that we found earlier, but we have
here a Yang--Mills sector and a matter sector. The latter can be
understood  as the  dimensional reduction  to two dimensions of, for
instance,   the eight-dimensional   Yang--Mills TQFT, provided one
adjusts the number of scalars $X$ and $Y$.

The     gauge symmetry is topological for all fields $A$, $X$ and
$Y$, and  is  made of arbitrary shifts  defined modulo gauge
transformations. The way
$\v$ transforms is interesting: it undergoes ordinary gauge rotations,
and  also rotates proportionally to the parameters of the shift
symmetry for $X$ and $Y$, according to:
\eqn\brtsA{\eqalign{  s \v_{{  }}&=-[c,\v] +[ U_{ 0 }^1, X_{}] +[ V_{ 
0}^1,  Y_{}] . }}  This forbids adding to the Lagrangian a $\v$ dependent
potential, except in the commuting limit, with a complex $\v$, in which
case we may have a Higgs potential.

 By applying the Batalin--Vilkoviski procedure, the  antifield-dependent
terms allow us to   determine   a fully gauge-fixed Lagrangian. One
can chose a gauge that eliminates half of the string coordinates, while,
for the remaining ones, one chooses  self-dual type ones,
$D_A X =*(D_A Y)+\ldots$, i.e., modified holomorphicity conditions.
Here, the 
$\ldots$ stand for commutators, which may  come
from    dimensional reduction
from eight to two dimensions.

We now investigate
the question of having more refined gauge functions,
for obtaining 
theories depending   on
$X$ or $Y$, with duality transformations between both formulations.

We introduce arbitrary
functions,
${\cal F}(\W_2, \U_1, \V_1) $,  and generalize  the   Lagrangian \IUV\  into:
\eqn\IUVd{\eqalign{\L_2 =   {\Tr} \ \Big(\ &
\U_{1} \w \V_{1}\ +\ 
\U_{1} \w  \DAT \X  \ +\ 
\V_{1}  \w \DAT \Y_{ }
 \cr &
+\B_{ } \w \W_2 
+  \B_{ } \w \FAT
+ \W_2 \w 
(\DAT \Z_{ -1} +[\X ,\Y_{ }])+ 
{\cal F}(\W_2, \U_1, \V_1) \ \Big)\ 
\Big| ^0_2 . \cr}}
If  we write that 
${\cal F}  =\tau \W_2+\ldots$, we see that  the remaining term at the
classical   level, after the integration over
$W_2$, is:
\eqn\IUVd{\eqalign{   {\Tr}  \big(  \tau F_A\big )
    }}

This is one way of understanding that we  are considering a topological
Yang--Mills theory in two dimensions. We will shortly see this
  from another point of view.

In the scalar sector, we can gauge-fix the fields in various ways, and
this provides different formulations. The strength of the coupling of
the resulting sigma models is  determined by  the derivatives  of
${\cal F}$ with respect to  $ \U_1   $ and $ \V_1   $.

Let us be slightly more    specific. Consider the following part of  the
classical Lagrangian:
 \eqn\IUVd{\eqalign{  {\Tr} \big( 
 \v_{ } \w  W_2 
+  \v_{ } \w \FA 
+  W_2 \w [ X , Y_{ }] \big) . }}
The classical two-form $W_2=dz \w d\bar z W_{z\bar z}$ determines its dual, the scalar 
$W=\e^{z\bar z} W_{z\bar z}$.

The first way of fixing the gauge, using the topological symmetry, 
is to set
$W_{z\bar z}=0$. This gives the ordinary topological gauge-invariant
Lagrangian:
 \eqn\IUVd{\eqalign{  {\Tr}  \big(
   \v_{ } \w \FA  \big)
  . }}
 Of course the other gauge symmetry in $A$
 must be used to gauge-fix this well-known two-dimensional Lagrangian,
that one sometimes uses to define the  topological two-dimensional
Yang--Mills theory. Then, the equations of motion imply that 
$\v$ must be constant, which gives back 
\IUVd.

\def\z{{\bar z}}

We could have first eliminated $\v$ and $W$ by their equations
of motion. Then, using the topological symmetry on $A$, with gauge
functions $F_{z\z}$ and
$\pa\cdot A$, we would have again directly obtained the two-dimensional
Yang--Mills TQFT, with its BRST-invariant gauge-fixing.

The other way  of expressing the theory, in 
a dual
formulation,  is to gauge-fix
$A$ as follows:
 \eqn\IUVdd{\eqalign{ 
A_z=g^{-1}\partial_{ z } g \cr 
A_\z=h^{-1}\partial_{ \z }h .  }}
The gauge group elements $g$ and $h$ are defined from the Lie
algebra-valued scalars  $W$ and $\v$  as:
\eqn\gh{\eqalign{ 
 g= \exp (\v+W) \quad\quad  
 h= \exp (\v-W) . }}
  This gauge is obtained by adding to the Lagrangian the   BRST-exact term 
$s\big ( 
 \bar\Psi_\z( A_z-g^{-1}\partial_{ z }g)
+\bar\Psi_ z (A_\z-h^{-1}\partial_{ \z }h ) \big)$.
The   bosonic  part of the Lagrangian has a   simple
expression if one restricts $W$ and $\v$ in 
the Cartan Lie algebra:  
\eqn\IUVd{\eqalign{  {\Tr} 
\big ( 
 \v_{ } W_{z\bar z}
+  \v_{ } \Delta W_{z\bar z}
+  W_{z\bar z}[ X , Y_{ }] \big) . }}
The BRST invariance implies the following ghost terms:
\eqn\IUVdd{\eqalign{ {\Tr} \ \Big(
\bar\Psi_z\big(\Psi_z -s (g^{-1}\partial_{ z } g)\big) + 
\bar\Psi_\z\big(\Psi_\z -s (h^{-1}\partial_{ \z } h)\big)
\ \Big)
}}

Another gauge-fixing term for the topological ghost is necessary:
\eqn\IUVddd{\eqalign{   s {\Tr} \ \Big(
 \Phi^{-2}_0 (D_z\Psi_\z +D_\z\Psi_ z) \Big ).  }}
We have  a massive fermionic Lagrangian, which is the supersymmetric
counterpart of 
\IUVd. The   term in \IUVddd\ ensures the inversibility of the fermionic
propagator, as well as the propagation of the  ghost   of ghost 
$\Phi^{ 2}_0$ and its antighost $\Phi^{-2}_0$.
Moreover a term    $s(\bar c \Phi^{0}_0)$ must be added to algebraically 
eliminate the n  unnecessary antighost sector. 
Of course,  terms of the form ${\cal F}(\W)^0_2$ can be added. Then,  
non-trivial interactions occurs between $W$ and the field $\Phi^2_0$.
We  also have the option of having 
a Higgs potential, by having the choice   ${\cal F}(\W, \v)^0_2$ in the
commuting limit, with a rotational symmetry between
$W$ and $\v$.

We actually  see  that the  two-form $W$ and the  zero-form $\v$,
 which
we   introduced  from  the  general considerations as  in \kyoto, have
a natural interpretation: they  can be used to express the
decomposition of  the gauge field in longitudinal and transverse parts,
which are specific to two dimensions, and the duality transformation
exchanges 
$W$ and  $\v$.
In the gauge $W=0$, $\v$ has also the interpretation of the 
Lagrange multiplier of the vanishing curvature condition, which is
usually interpreted as the characteristic of a Yang--MIlls TQFT in two
dimensions.

\vskip 0.5cm

\centerline{\bf Acknowledgements}

We are grateful to L. Alvarez-Gaume. W. Lerche,  N. Nekrasov, E. Rabinovici, I. M. Singer,
S. Stieberger, E. Verlinde and  N. Warner  
  for discussions.
  The research of
S.~Shatashvili is supported 
by DOE grant DE-FG02-92ER40704, by NSF CAREER award, by
OJI award from DOE and by Alfred P.~Sloan foundation.

 \listrefs

\bye